\documentclass[12pt]{article}
\usepackage{amsmath}
\usepackage{graphicx}
\usepackage{natbib}
\usepackage{url} % not crucial - just used below for the URL

\newcommand{\blind}{1}

\usepackage[table]{xcolor}

\usepackage[utf8]{inputenc}
\usepackage[russian,english]{babel}

\usepackage{amsthm, amssymb, dsfont, multirow}
\newcommand{\xvec}{\boldsymbol}
\newcommand{\xmat}{\mathbf}
\newcommand{\xset}{\mathds}
\newcommand{\xcal}{\mathcal}
\newcommand{\indicator}{1\!\!1}
\DeclareMathOperator*{\argmin}{arg\,min}

\newtheorem{proposition}{Proposition}

\usepackage{tikz}

\begin{document}

\def\spacingset#1{\renewcommand{\baselinestretch}%
{#1}\small\normalsize} \spacingset{1}

%%%%%%%%%%%%%%%%%%%%%%%%%%%%%%%%%%%%%%%%%%%%%%%%%%%%%%%%%%%%%%%%%%%%%%%%%%%%%%

\if1\blind
{
  \title{\bf Estimation of the Spatial Weighting Matrix for Spatiotemporal Data under the Presence of Structural Breaks}
  \author{Philipp Otto \footnote{Corresponding author (email: otto@ikg.uni-hannover.de)}\\
    Leibniz University Hannover, Germany\\
    and \\
    Rick Steinert \\
    European University Viadrina, Frankfurt (Oder), Germany}
  \maketitle
} \fi

\if0\blind
{
  \bigskip
  \bigskip
  \bigskip
  \begin{center}
    {\LARGE\bf Estimation of the Spatial Weighting Matrix for Spatiotemporal Data under the Presence of Structural Breaks}
\end{center}
  \medskip
} \fi

\bigskip
\begin{abstract}
In this paper, we propose a two-stage LASSO estimation approach for the estimation of a full spatial weights matrix of spatiotemporal autoregressive models. In addition, we allow for an unknown number of structural breaks in the local means of each spatial location. These locally varying mean levels, however, can easily be mistaken as spatial dependence and vice versa. Thus, the proposed approach jointly estimates the spatial dependence, all structural breaks, and the local mean levels. For selection of the penalty parameter, we propose a completely new selection criterion based on the distance between the empirical spatial autocorrelation and the spatial dependence estimated in the model. Through simulation studies, we will show the finite-sample performance of the estimators and provide practical guidance as to when the approach could be applied. Finally, the method will be illustrated by an empirical example of {intra-city} monthly real-estate prices in Berlin between 1995 and 2014. The spatial units will be defined by the respective {postal} codes. The new approach allows us to estimate local mean levels and quantify the deviation of the observed prices from these levels due to spatial spillover effects. {In doing so, the entire spatial dependence structure is estimated on a data-driven basis.}
\end{abstract}

\noindent%
{\it Keywords:}  LASSO estimation, spatiotemporal autoregressive models, spatial weights matrix, structural breaks, real-estate market.
\vfill

% \newpage
\spacingset{1.5} % DON'T change the spacing!

\section{Introduction}\label{sec:introduction}

Spatial autoregressive models, along with conditional autoregressive models, require a definition of a suitable spatial dependence structure via the so-called spatial weights matrix (see e.g., \citealt{Hoef18,Elhorst10}{, \citealt{Lesage09})}. This matrix, like an adjacency matrix in graphical models, defines how the locations are connected, thereby defining which locations have a possibility of being dependent, along with to what extent. The spatial autoregressive term is then classically modeled as a multiple of the product of this predefined weighting matrix and the vector of observations. Indeed, this also means that estimated spatial autoregressive coefficient depends on the choice of this matrix. Thus, all coefficients, as well as inference on these parameters, should always be done conditionally, based on the definition of the weighting scheme. {Otherwise, wrong conclusions would be drawn. This sensitivity on the spatial weights led to a lot of criticism on these models (see also \citealt{lesage2014biggest}). In practice, the weights must be predefined or selected from a set of candidate weighting schemes.} This approach is, however, quite unsatisfactory, as the underlying spatial dependence structure is unknown in almost all cases. Thus, several papers have analyzed the impact of misspecified weighting matrices, e.g., \cite{Stakhovych09}. Moreover, the entire matrix of spatial weights cannot be estimated by classical approaches such as the maximum likelihood method (cf. \citealt{Lee04}), generalized method of moments (cf. \citealt{Kelejian99}), or least squares procedures (cf. \citealt{Kelejian98a}), due to the curse of dimensionality and endogeneity. Moreover, even if the sample size is large enough to estimate the weights, the weights matrix might not be identifiable without further assumptions. For instance, \cite{Lam16,Lam19} employ a certain sparsity structure and \cite{Bhattacharjee13} assume symmetric weights with constant {degree of} spatial dependence across space.\footnote{{More precisely, the autoregressive parameter is assumed to be constant across all spatial units.}} Furthermore, classical ordinary least squares estimators are inconsistent in the presence of spatial dependence. In addition, it is often observed that there are anisotropic dependencies with unobserved exogenous effects (e.g., \citealt{Deng08,Fasso13}), such that the classical parametric approach might cause issues due to misidentified dependencies.

In this paper, we propose a penalized regression approach to estimate the entire spatial weighting matrix of a dependent spatiotemporal process, as well as an unknown number of change points, which may occur at different time points depending on the location. The current literature contains only a few papers that propose the LASSO-type estimation procedures for similar spatiotemporal models. Thus, the majority of studies that apply methods of spatial econometrics still assume that the spatial weights matrix would be known a priori.

Pioneering works on estimating spatial dependence structures in terms of weighting matrices in spatial autoregressive models can be found in \cite{Bhattacharjee13,Lam13,Ahrens15}. Whereas \cite{Bhattacharjee13} propose a two-stage estimation procedure based on a spatial autocovariance matrix estimated in the first stage, \cite{Lam13,Ahrens15} consider (adaptive) LASSO-type estimation procedures. Moreover, \cite{Lam16} focus on a spatiotemporal model setup with exogenous regressors, in which the matrix of spatial weights has a block diagonal structure.

Unlike these approaches, we account for structural breaks, which may occur in the temporal dimension. This leads to a flexible modeling approach, but a high number of parameters needed to be estimated. Thus, we propose a two-stage constrained LASSO estimation approach for the simultaneous estimation of the spatial dependence and change points in the local means. It is important to note that these change points are not necessarily the same for all spatial locations. In addition, {due to the spatial dependence,} a change in one location would affect {the neighboring or even} all of the other locations as well. {Moreover, local mean changes in multiple adjacent locations likely interfere with each other and, thus, get strengthened or diminished. Hence}, the spatial dependence can easily be mistaken as break points, and vice versa, and a major issue is to distinguish between the changes in the observed mean level that are due to the spatial dependence {(i.e., changes that spill-over from neighboring locations)} or due to a structural break. This distinction gets even more complicated, if there are breaks in multiple locations at the same time point. In contrast, \cite{Angulo17} consider structural breaks in the spatial dependence structure, which is assumed to have a certain structure.

One fundamental assumption in spatial statistics originates in the first law of geography, namely ``everything is related to everything else, but near things are more related than distant things'' (\citealt{Tobler70}). However, in our approach the estimated weights are not necessarily due to spatial proximity, but might also be caused by other physical drivers or dependencies. As a consequence, we can also detect dependencies {going} beyond spatial proximity, although it is designed for so-called dynamic spatial panel data models including a spatial autoregressive term and an unknown spatial weights matrix. In general, our approach, as well as the one presented in \cite{Lam16}, are related to the estimation of (high-dimensional) covariance matrices (e.g. \citealt{Cai16,Bodnar16}). {Hence, the geographical meaning and interpretation of the weights must be done after the estimation, e.g., via graphical inspection on a map.}

If the model errors are assumed to be independent and normally distributed, the covariance matrix of the observed process can directly be computed from the estimated spatial weights matrix. {These spatial weights also directly provide information about the conditional independence between spatial locations.\footnote{{All zero entries in  $(\xmat{I} - \xmat{W}) (\xmat{I} - \xmat{W}')$ imply conditional independence between the corresponding location pairs. Here, $\xmat{W}$ denotes the spatial weights matrix and $\xmat{I}$ is an identity matrix of the same dimension.}}} Alternatively, {the} covariance matrix can be modeled directly using parametric or nonparametric spatial covariance functions. This is the usual approach in geostatistics. In this field, the concepts of (weak) spatial stationarity and isotropy are important. In the first case, the covariance between two observations is only a function of the difference of their coordinates and in the latter case this dependence is even independent of the orientation between the locations (cf. \citealt{Cressie11}). In contrast to spatial econometrics, there are many approaches in geostatistics, which do not (explicitly or implicitly) assume stationarity or isotropy (e.g. \citealt{Zhu10a,Porcu06,Fuentes02,Sampson92}). For instance, \cite{Schmidt03,Schmidt11} introduced the idea of a latent space approach, which can include covariates in the spatial covariance function. These approaches could also violate Tobler's first law, but do not have to. Moreover, there exists papers applying penalized estimators in geostatistics, e.g. \cite{Chu11} for variable selection or \cite{Krock20} for spatial kriging using graphical LASSO.

The remainder of the paper is organized as follows. The considered spatiotemporal autoregressive model is presented in the next section.
Subsequently, we propose a new two-stage LASSO approach in Section \ref{sec:LASSO}. To evaluate the performance of the proposed procedure, a series of Monte Carlo simulations will be presented in the ensuing section, followed by an illustration of the procedure via an empirical example of Berlin condominium prices from 1995-2014. Section \ref{sec:conclusion} concludes the paper.

% Van Hoef ergänzen

\section{Spatiotemporal Model}\label{sec:model}

We consider that the data is generated by a univariate spatiotemporal process {$\{Y_t(\xvec{s}): t \in \xcal{D}_t, \xvec{s} \in \xcal{D}_s\}$}, where {$\xcal{D}_t$} is the temporal domain and {$\xcal{D}_s$} is the spatial domain of the process. {Regarding the temporal domain, we consider discrete time points which are typically regularly spaced, but could also be irregularly spaced.} The most common method would be to consider {$\xcal{D}_t$} as a subset of all integers $\xset{Z}$. The spatial domain $\xcal{D}_s$ can either be defined as a {discrete} subset of the $d$-dimensional real numbers, $\xcal{D}_s \subseteq \xset{R}^d$, or as a {discrete} subset of the $d$-dimensional integers $\xset{Z}^d$. For the first case, {geostatistical} and marked point processes are covered,{\footnote{{Note that our focus is not on modelling the (random) locations of the process, but primarily on the dependence in the marks. Moreover, we later assume that the points do not change across time.}}} e.g., data of ground-level measurement stations ($d = 2$), {{areal} data {on irregular polygons}, like} financial or economic data of certain regions, states, municipalities ($d = 2$), or atmospheric and satellite data ($d = 3$). In contrast, the second case covers all kinds of spatial lattice data, such as images ($d = 2$), CT scans ($d = 3$) or raster data ($d = 2$). Moreover, the locations $\xvec{s}$ can also be understood as vectors of $d$ characteristics that are associated with $Y$. For economic studies, such as those of regional labor markets, these characteristics could comprise the region's population, gross domestic product, area, and other factors. It is important to note that our methods cannot be used for spatial kriging at unknown locations as it is often done in geostatistics. Furthermore, temporal forecast are not in the focus of our paper, because we assume that there are temporal structural breaks.

\subsection{{Model Definition}}

{We} assume that the process {is observed} for $t = 1, \ldots, T$ time points at a constant set of {spatial} locations $\{\xvec{s}_1, \ldots, \xvec{s}_n\}$. For convenience of notation, we denote the time point as an index in the following sections. {The} vectors of observations of the spatiotemporal random  process $\{\xvec{Y}_{t \cdot} = (Y_t(\xvec{s}_1), \ldots, Y_t(\xvec{s}_n))': t = 1, \ldots, T\}$ are given by $\xvec{y}_{t \cdot} = (y_t(\xvec{s}_1), \ldots, y_t(\xvec{s}_n))'$ for $t = 1, \ldots, T$. Furthermore, we introduce the $T \times n$-dimensional matrix $\xmat{\Upsilon} = (Y_{t}(\xvec{s}_i))_{t = 1, \ldots, T; i = 1, \ldots, n}$ {representing the panel format}. 

{We consider that the process has an autoregressive dependence structure across space and that the location-specific mean level of the process might change at an unknown number of time points.} These breaks may occur at different time points for each location $i$ and they can be of a different magnitude. Thus, at time point $t$, the considered spatial autoregressive model can be specified as
\begin{equation}\label{eq:model}
\xvec{Y}_{t \cdot} = \xmat{W}\xvec{Y}_{t \cdot} + \xvec{a}_t + \xvec{\varepsilon}_t \, ,
\end{equation}
where the vector $\xvec{\varepsilon}_t$ denotes an independent and identically, normally distributed random error with zero mean and constant variance across space and time. The {location-specific, or} local mean levels are defined by $\xvec{a}_t = (a_{t,1}, \ldots, a_{t,n})'$ {and $\xmat{W}$ denotes the unknown spatial interaction matrix (i.e., the spatial weights matrix)}. It is important to note that we {refer to} $\xvec{a}_t$ {as \emph{local}} mean levels, because they do not correspond to the true process mean {$\xvec{\alpha}_t = E(\xvec{Y}_{t\cdot}) = (\xmat{I} - \xmat{W})^{-1} \xvec{a}_t$}. {However, any change in the process' mean $\xvec{\alpha}_t$ can only be due to changes in $\xvec{a}_t$, i.e., the parameters to be estimated.} 

To illustrate the interdependence {$\xvec{a}_t$ and $\xvec{\alpha}_t$}, we depict a simulated $3 \times 3$ random field as time-series plots {with $T = 200$ time points} in Figure \ref{fig:example}. The spatial weights has been chosen as first-order Queen's contiguity matrix, where each row sums to $0.9$. That is, each of the 9 locations equally influence its direct neighbors.\footnote{{Note that this coincides with a classical spatial autoregressive model with $\rho = 0.9$ and a row-standardized Queen's contiguity matrix, which is visualized in the appendix for reference.}} There are two change points at $t = 50$ and $t = 100$ where the local mean level {of the first location} $a_{t,1}$ {changes}. {The local mean levels $\xvec{a}_t$ are drawn as a solid red line. Due} to the spatial dependence, the {process' mean} level in the first location is higher than the red line, and the change ripples out to other locations, as can be seen by the overall mean level {$\xvec{\alpha}_t$} depicted by the dashed {blue} curve.

\begin{figure}
  \centering
  \input{Example_3_3.tex}
  \caption{Simulated spatial panel with $n = 9$ locations and $T = 200$ points of time. The true {local mean} levels $\xvec{a}_t$ are plotted in red (solid line), whereas the true overall mean level $\xvec{\alpha}_t = (\xmat{I} - \xmat{W})^{-1} \xvec{a}_t$ is drawn as a blue curve (dashed line)}\label{fig:example}
\end{figure}

The spatial dependence is assumed to be defined by the $n \times n$-dimensional spatial {interaction} matrix {$\xmat{W} = (w_{ij})_{i,j = 1, \ldots, n}$}, which is constant over time. {Each element $w_{ij}$ of the matrix $\xmat{W}$ defines how the $i$-th location is influenced by the $j$-th location.} Moreover, the diagonal elements of the matrix are assumed to be zero in order to prevent observations from influencing themselves. In spatial econometrics, the classical approach would be to replace the unknown spatial dependence structure {$\xmat{W}$} with a linear combination $\rho \tilde{\xmat{W}}$, where it would be assumed that $\tilde{\xmat{W}}$ is a predefined, non-stochastic {$n \times n$} weighting matrix that describes the dependence {structure and $\rho$ is an unknown scalar parameter that has to be estimated}. In practice, however, the true underlying {weights} matrix $\tilde{\xmat{W}}$ cannot easily be assessed, and therefore has to be {selected from a set of candidates} by maximizing certain goodness-of-fit measures, such as the log-likelihood, in-sample fits, information criteria, or {other error measures obtained by} cross validation. {For instance, \cite{zhang2018spatial} showed that the true weighting matrix can consistently be selected by a Mallows type criterion, if the true one is under the candidate schemes. Moreover, even if the true weight matrix is not under the candidate schemes, the best fitting weight matrix in terms of minimal squared losses can be selected by this approach (see \citealt{zhang2018spatial}).} {It is also not directly possible to derive a suitable weighting matrix from the variogram or semivariogram, although these provide initial information about the distance decay of the dependencies (see \citealt{Cressie93,Cressie11}).} Other papers, such as \cite{Stakhovych09}, have aimed to assess the size of the effect of misspecified weighting matrices, as one might expect that missing spatial links could be compensated through other linkages.

{The overall mean $\xvec{\alpha}_t = (\xmat{I} - \xmat{W})^{-1} \xvec{a}_t$ shows that it is important that $(\xmat{I} - \xmat{W})$ is invertible for this class of models.} In general, the spatiotemporal autoregressive model is well-defined and stationary, if and only if the series
\begin{equation}\label{eq:series}
(\xmat{I}_n - \xmat{W})^{-1} = \xmat{I}_n + \xmat{W} + \xmat{W}^2 + \xmat{W}^3 + \cdots
\end{equation}
converges. This is the case, if and only if
\begin{equation*}
    \varrho(\xmat{W}) < 1 \, ,
\end{equation*}
where $\varrho(\cdot)$ is the spectral radius. Moreover, if $\varrho(\xmat{W}) < 1$, then there exists a norm $|| \cdot ||$, such that the series in \eqref{eq:series} converges, according to Gelfand's formula (see \citealt{Gelfand41,Kozyakin09}). % Gelfand41 % Gelfand's theorem, von Neumann’s theorem
In spatial {econometrics}, the weighting matrix $\tilde{\xmat{W}}$ is often row-standardized, such that $|| \tilde{\xmat{W}} ||_{\infty} = 1$ and $||\rho \tilde{\xmat{W}}||_{\infty} < 1$ for all $|\rho| < 1$. The row-standardization also implies that the elements $\tilde{w}_{ij}$ are less than or equal to one and that all absolute row sums would be equal to one. Our proposed constrained LASSO approach {ensures that $\varrho(\xmat{W}) < 1$ in terms of the maximum row sum being smaller than one. The same condition is also considered in} \cite{Lam16}{, namely $||\xmat{W}||_{\infty} < 1$. Note that this is only a necessary condition.}

Furthermore, {the set of changes in the local mean levels of the $i$-th location} is given by 
\begin{equation}
	{\xcal{T}_i = \{\tau : a_{\tau, i} \neq a_{\tau-1,i} \} \, .}
\end{equation}
Since the mean level can change at $(T-1)$ different points of time for all $n$ locations, there are $n(T-1)$ possible mean changes. However, we assume that there is a sparse representation of all mean changes {$\xvec{a}_t$. That is, they remain constant for certain periods. When a change point occurred, the local mean levels can change at some locations (i.e., some entries of $\xvec{a}_t$ change) or at all locations (i.e., all entries of $\xvec{a}_t$ change).} These locally varying mean levels are important to identify the spatial dependence. {In contrast,} if $\xvec{a}_t$ would be zero for all time points and locations, there would be infinitely many solutions to reconstruct the spatial dependence (without further assumptions, like symmetry) {and} further exogenous regressors or instrumental variables would be needed (cf. \citealt{Lam13,Merk19_arxiv}).

Due to the very flexible structure of $\xvec{a}_t$ allowing for different mean levels at all locations and time points, the mean influence of omitted exogenous regressors would be intercepted by $\xvec{a}_t$. {Alternatively,} one could replace the local mean levels by a combination of the structural break model and a regressive term $\xvec{X}_t \xvec{\beta}$ in order to incorporate exogenous effects, i.e., the model equation would become
 \begin{equation}
	\xvec{Y}_{t \cdot} = \xmat{W}\xvec{Y}_{t \cdot} + \xvec{a}_t + \xvec{X}_t \xvec{\beta} + \xvec{\varepsilon}_t \, .
 \end{equation}
 Moreover, $\xvec{a}_t$ can be interpreted as common shocks. Further possible extensions and adaptations are discussed in Section \ref{sec:conclusion}.

\subsection{{Number of Parameters and Identification}}

{Because of the spatial interactions,} all mean changes and the spatial dependence structure must be estimated simultaneously, {such that} there are $n(n + T - 1)$ {unknown} parameters to be estimated. That is, the number of parameters is much larger than the number of observations $nT$, which leads to the system of equations being overdetermined. In other words, there is no unique solution to the problem for the $nT$ observations, unless the number of parameters is reduced. Moreover, even if {the} parameter space is reduced (i.e., the number of parameters is equal to or less than the size of the sample), the parameter sets leading to $\xvec{Y}_{t \cdot}$ might be observationally equivalent. This is a common but yet often overlooked problem in spatial econometrics (see, for instance, \citealt{Manski93,Gibbons12}). The case where $\xmat{W}$ and $\xvec{a}_t$ have to be estimated together is specifically prone to this identification issue. Thus, we briefly discuss under which circumstances the model parameters are identifiable.

{Recall, instead of assuming a known weight structure, we aim to estimate the entire matrix $\xmat{W}$ through a penalized regression. Because the mean levels $\xvec{a}_t$ are simultaneously estimated, we have to distinguish between changes in $\xvec{\alpha}_t$ that are due to changes in $\xvec{a}_t$ or due to the spatial interactions implied by $\xmat{W}$. In other words, even when ignoring the spatial dependence (i.e., $\xmat{W}$ is a zero matrix), the overall mean $\xvec{\alpha}_t$ could be obtained by a different set of local mean levels $\xvec{a}_t$ that mimic the spatial interaction by an increased number of change points for each location. To analyze this interaction, we rewrite the} model {in its reduced form} as
\begin{equation}\label{eq:model2}
\xvec{Y}_{t \cdot} = (\xmat{I}_n - \xmat{W})^{-1}\xvec{a}_t + (\xmat{I}_n - \xmat{W})^{-1}\xvec{\varepsilon}_t \, ,
\end{equation}
where $\xmat{I}_n$ stands for the $n$-dimensional identity matrix. Thus, a change $a_{t,i}$ at location $i$ does not only influence the observed value at location $i$, but also {other} locations via the spatial dependence that is implied by $\xmat{W}$. {Below, let} $\xmat{S} = (s_{ij})_{i,j = 1, \ldots, n}$ be the inverse $(\xmat{I}_n - \xmat{W})^{-1}$. Having a closer look at the expected value
\begin{align*}
E(Y_{t}(\xvec{s}_i)) &= \alpha_{t,i} = \sum_{j = 1}^{n} s_{ij} a_{t,j} \, ,
\end{align*}
we see that {both effects can easily be separated} if either the spatial dependence or the mean levels are known. However, if both parameters have to be estimated, one can only distinguish between $s_{ij}$ and $a_{t,j}$ {because} the spatial dependence structure is constant over time, whereas the mean levels depend on $t$. {Furthermore, \eqref{eq:model2} shows that the error term $(\xmat{I}_n - \xmat{W})^{-1}\xvec{\varepsilon}_t$ is spatially correlated and the weight matrix $\xmat{W}$ is the only source of cross-correlation in our model}. The random errors {$\xvec{\varepsilon}_t$} are assumed to be independent across space and time. {Thus,} any additional source of cross-correlation, which is not explicitly modeled, would be associated with the weight matrix $\xmat{W}$. {Therefore, the (spatial) independence assumption of the errors is crucial in our model.} {Moreover,} the process is uncorrelated over time, {such that} mean changes are {instantly} reflected into neighboring locations. On the contrary, if there are temporal dependencies, we would observe faded spill-over effects of the mean changes. Even though temporal correlation is an important point for future research, we initially only focus on the case without (autoregressive) dependence over time in this paper. 

{First,} we start by analyzing under which conditions the model parameters can be identified for a simple example. We therefore focus on the case with at least two time points having a constant mean level and one change point. Let $\vartheta = \{\xmat{W}, \xvec{a}_1, \ldots, \xvec{a}_T\}$ be the set of true parameters and $\tilde{\vartheta} = \{\tilde{\xmat{W}}, \tilde{\xvec{a}}_1, \ldots, \tilde{\xvec{a}}_T\}$ be a second, observationally equivalent set of parameters. Then,
\begin{equation*}
  E(\xvec{Y}_{t \cdot}) = (\xmat{I} - \xmat{W})^{-1} \xvec{a}_t = (\xmat{I} - \tilde{\xmat{W}})^{-1} \tilde{\xvec{a}}_t \qquad \text{for all $t = 1, \ldots, T$.}
\end{equation*}
To ensure that $\xmat{W}$ can be identified, a certain number of change points must be observed, but also locally constant mean levels. Suppose that there exists $t_1$, $t_2$, and $t_3$, such that $\xvec{a}_{t_1} = \xvec{a}_{t_2} \neq \xvec{a}_{t_3}$ and all of them are different from the zero vector. If $\tilde{\vartheta}$ is observationally equivalent, then
\begin{eqnarray*}
  (\xmat{I} - \xmat{W})^{-1} \xvec{a}_{t_v} - (\xmat{I} - \tilde{\xmat{W}})^{-1} \tilde{\xvec{a}}_{t_v} & = & \xvec{0} \qquad \text{for $v = 1, 2, 3$, and}\\
  (\xmat{I} - \xmat{W})^{-1} \underbrace{(\xvec{a}_{t_1} - \xvec{a}_{t_3})}_{\neq \xvec{0}}  - (\xmat{I} - \tilde{\xmat{W}})^{-1} (\tilde{\xvec{a}}_{t_1} - \tilde{\xvec{a}}_{t_3}) & = & \xvec{0} \, , \\
    (\xmat{I} - \xmat{W})^{-1} \underbrace{(\xvec{a}_{t_1} - \xvec{a}_{t_2})}_{= \xvec{0}}  - (\xmat{I} - \tilde{\xmat{W}})^{-1} (\tilde{\xvec{a}}_{t_1} - \tilde{\xvec{a}}_{t_2}) & = & \xvec{0} \,.
\end{eqnarray*}
There is no solution of this system of equations, unless $\xmat{W} = \tilde{\xmat{W}}$. Then, $\vartheta = \tilde{\vartheta}$. {To ensure the identifiability of $\xmat{W}$, there must be sufficient variation {(i.e., sufficiently many changes)} in the local mean levels {$\xvec{a}_t$} as we will show in the following proposition. For this reason,} let $\xvec{\iota}_{ij}$ be {an $n \times n$} matrix with the $(i,j)$-th entry being one and {all remaining entries are zero. Furthermore, we define} $\xmat{A}_{\mathcal{X}}$ {as a} submatrix of $\xmat{A}$ {with only the columns that are included in the index set} $\mathcal{X}$. Below, these columns are given by the true active sets of the mean levels and the spatial dependence, which are given by $\mathcal{A}_a$ and $\mathcal{A}_w$, respectively. That is, these two active sets consist of the indices of the non-zero parameters{, i.e., the (non-zero) changes in local mean levels, $a_{t,i} - a_{t-1,i}$, and the non-zero weights $w_{ij}$}. In addition, let {$\xvec{a}$ (without index)} be the stacked $nT$-dimensional vector of all local mean levels, i.e., {$\xvec{a} = (\xvec{a}_{1}, \ldots, \xvec{a}_{T})'$.}

\begin{proposition}\label{prop:1}
  Suppose that $\{\xvec{\varepsilon}_t\}$ is an independent and identically distributed zero mean process and that $\varrho(\xmat{W}) < 1$. Moreover, {$a_{\tau,i} \neq a_{\tau - 1,i}$ for all $(\tau, i)' \in \mathcal{A}_a$} and $\xmat{W}$ has nonzero entries for all $(i,j)' \in \mathcal{A}_w$. Let
  \begin{equation}
  \tilde{\xmat{H}} = \left[\left(\xmat{I}_T \otimes \xmat{S}\right)_{\mathcal{A}_a}, ((\xmat{I}_T \otimes \xmat{S} \xvec{\iota}_{ij} \xmat{S})\xvec{a})_{(i,j)' \in \mathcal{A}_w}\right].
  \end{equation}
   If $|\mathcal{A}_a \cup \mathcal{A}_w| \leq nT$ and the nullity of $\tilde{\xmat{H}}$ is zero, then the parameters $\vartheta$ of the model given by \eqref{eq:model} are locally identifiable in an open neighborhood of $\vartheta$.
\end{proposition}

To put it briefly, the true number of change points, as well as the number of non-zero entries in $\xmat{W}$, must be smaller than $nT$ and there should not be linear dependencies in $\tilde{\xmat{H}}$. {The first condition ensures that the model is not high-dimensional (i.e., the number of non-zero parameters is smaller than the number of observations). On the contrary, in the high-dimensional setting, there are always multiple observationally equivalent models. We will focus on this case in the {last} paragraph {of this section}. The second condition on the linear dependencies in $\tilde{\xmat{H}}$ is more difficult to check. This matrix $\tilde{\xmat{H}}$ must have full rank that coincides with the number of true non-zero parameters (i.e., the number of columns of $\tilde{\xmat{H}}$). It ensures that there is ``sufficient'' variation in the local mean levels to identify $\xmat{W}$. Note that the example in Figure \ref{fig:example} does not fulfil the criterion in Proposition \ref{prop:1}.} 

To strengthen the understanding of this sufficiency, we briefly discuss a numerical example. Assume a relatively short time frame with $T = 50$ and $n = 16$ locations {arranged in a 4 by 4 lattice}, which are connected by a Queen's contiguity matrix. {Consequently, there are $84$ parameters{\footnote{The 4 interior lattice points have 8 neighbors each ($4 \cdot 8$ parameters); the non-corner lattice points on the edges have either 5 neighbors ($8 \cdot 5$ parameters), and the four corner points have 3 neighbors ($4 \cdot 3$ parameters), resulting in a total of 84 parameters.}} in the true active set $\mathcal{A}_w$ and $((\xmat{I}_T \otimes \xmat{S} \xvec{\iota}_{ij} \xmat{S})\xvec{a}_L)_{(i,j)' \in \mathcal{A}_w}$ has 84 columns corresponding to the true non-zero weights {$w_{ij}$ for} $(i,j) \in \mathcal{A}_w$.} We then compute the rank of $\tilde{\xmat{H}}$ for varying numbers of change points {$\kappa$}. For each case, we randomly sample {the time points and the magnitude of the mean changes}. {That is, there are $n (\kappa+1)$ parameters in the true active set $\mathcal{A}_a$, such that the matrix $\tilde{\xmat{H}}$ has $nT = 800$ rows and $84 + n (\kappa+1)$ columns.} The results {of this numerical study} are depicted in Figure \ref{fig:uniqueness}. In this graph, there are two dashed lines {visible}, which represent the necessary rank of $\tilde{\xmat{H}}$ for which the problem is identifiable. The upper dashed line depicts the aforementioned condition of the number of parameters being less than $nT=800$, while diagonal one ensures that the nullity of $\tilde{\xmat{H}}$ is zero (i.e., $\tilde{\xmat{H}}$ has full rank). All identifiable solutions must be found exactly on the diagonal dashed line and can not exceed the horizontal dashed line. The solid black line represents the rank of $\tilde{\xmat{H}}$ depending on the number of change points. In our example, all parameters were uniquely identifiable if there were more than six and less than 41 breakpoints. Each of these settings is depicted as a vertical dotted line. Moreover, it is worth noting that for more than 45 breakpoints the problems becomes high-dimensional, i.e., the number of parameters exceeds the number of observations. 

\begin{figure}
  \centering
  \input{Identifiability.tex}
  \caption{Rank of $\tilde{\xmat{H}}$ with respect to a different number of {local mean levels $\kappa + 1$}. The vertical, gray dashed line indicate the cases, where all parameters were uniquely identifiable.}\label{fig:uniqueness}
\end{figure}

In practice, {the number of parameters always exceeds sample size}, which is a typical issue of high-dimensional statistics. To solve that, we propose a two-stage adaptive LASSO approach. This approach is tailor-made for such situations, since all solutions for each choice of the penalty parameter will lead to the same fitted values $\hat{Y}_t(\xvec{s}_i)$ (cf. \citealt{Tibshirani12}). That is, the LASSO solution path always produce a unique solution of the fitted values -- even if there are too many change points or too many non-zero weights (cf. \citealt{Tibshirani12,Tibshirani13,Ali19}). For a detailed analysis of the uniqueness specifically for the adaptive LASSO, we refer to \cite{Zhou09arxiv_adaptive}. {Summarizing this section, we can state that we are generally in a high-dimensional setting, where the number of parameters is larger than the number of observations. In this high-dimensional case, the interpretation of the estimated weights must be done with caution, because there is generally no unique solution of the model parameters and only the fitted values $\hat{Y}_t(\xvec{s}_i)$ are unique. However, if the true number of non-zero spatial weights and local mean changes does not exceed the sample size, we can obtain (local) identifiability of the parameters. In a simulation, we showed that the necessary rank condition was fulfilled in the majority of cases for sparse Queen's contiguity matrix with at least six changes in the local mean levels.}

\section{Two-Stage Lasso Estimation}\label{sec:LASSO}

The motivation behind the two-stage approach is to initially separate the change point detection problem from the spatial dependence. Thus, we begin by estimating a set of candidate change points {$\xcal{T}_i^{(1)}$} for each location {$\xvec{s}_i \in \{\xvec{s}_1, \ldots, \xvec{s}_n\}$}. In the first stage, we only consider the univariate time series $\xvec{Y}_{\cdot i} = (Y_1(\xvec{s}_i), \ldots, Y_T(\xvec{s}_i))'$, i.e., $\xvec{Y}_{\cdot i}$ is the $i$-th column of $\xmat{\Upsilon}$. {That is, we essentially estimate the overall mean levels $\xvec{\alpha}_t$ based on the individual time series.} Even ignoring the spatial dependence, {the set of all change points} {$\{\xcal{T}_i^{(1)} : i = 1, \ldots, n\}$} can {therefore} be estimated without a loss of information, as the spatial dependence is not time-varying and there is no additional source of temporal dependence. In addition, the number of estimated parameters in the full model is reduced if there are fewer candidate change points than time points $T$, i.e., the cumulated number of local change points {$\left|\bigcup_{i=1}^{n}\xcal{T}_i^{(1)}\right|$} is less than $T$, which makes the approach computationally feasible. These sets of candidate change points are then passed to the full model to estimate $\xvec{a}_{1}, ..., \xvec{a}_{T}$, and $\xmat{W}$, simultaneously. In the second stage, we suppose that for all locations {$\xvec{s}_i$} the mean level $a_{t+1,i}$ is equal to $a_{t, i}$, if {$t \notin \xcal{T}_i^{(1)}$}. As the set of all estimated change points {$\hat{\xcal{T}}_i = \{\tau : \hat{a}_{\tau,i} \neq \hat{a}_{\tau-1,i} \}$} in the second stage would always be a subset of {$\xcal{T}_i^{(1)}$}, i.e., {$\hat{\xcal{T}}_i \subseteq \xcal{T}_i^{(1)}$}, it is important that we do not underestimate the number of possible change points in the first stage, but yet eliminate enough to ensure the computational feasibility of the problem.

\subsection{First stage: candidate change points}

To find all candidate change points {$\xcal{T}_i^{(1)}$} for location $i$, consider the following linear models for $i = 1, \ldots, n$, where
\begin{equation*}
\xvec{Y}_{\cdot i} = \xmat{K}\tilde{\xvec{\alpha}}_i + \xvec{\varepsilon}_{i}
\end{equation*}
and $\xmat{K} = (\indicator_{\{j, \ldots, T\}}(i))_{i,j = 1, \ldots, T}$ is a $T \times T$ lower triangular matrix and $\indicator_A(x)$ denotes the indicator function on the set $A$. The vector of coefficients $\tilde{\xvec{\alpha}}_i = (\tilde{\alpha}_{t,i})_{t = 1, \ldots, T}$ is used to find the candidate change points (i.e., $\tilde{\xvec{\alpha}}_i$ are changes of the overall mean levels). {The vector $\xmat{K}\tilde{\xvec{\alpha}}_i$ coincides with the overall mean levels $\xvec{\alpha}_i = \sum_{j=1}^{n} s_{ij} a_{t,j}$ (i.e., including all spatial spillovers).}
However, this specific model setup leads to the case where the number of estimated parameters would be exactly the same as the number of observations $T$. Any solution using the full matrix $\xmat{K}$ will, therefore, exactly fit the model to the observed values (i.e., $\hat{Y}_t(\xvec{s}_i) = y_t(\xvec{s}_i)$ and $\varepsilon_t(\xvec{s}_i) = 0$ for all $i$ and $t$) and will not provide a meaningful interpretation. Consequently, we need to be able to set at least some of the elements of $\tilde{\xvec{\alpha}}_i$ to zero, leading to our  proposal for the use of a penalized regression approach.
Recent literature has shown that certain types of penalized regressions can be used {for change point detection}. For instance, \cite{chan2014group} used a group-LASSO approach to estimate the structural breaks in an autoregressive time series. In contrast, \cite{lee2018oracle} utilized a SCAD-type penalty to provide an estimator for change points, with the oracle property in quantile regression.

For our setup, we propose the use of the adaptive LASSO penalty, first introduced by \cite{zou2006adaptive}{, which is less biased for the true non-zero parameters compared to the classical LASSO.}
Moreover, and more importantly, the approach must ensure that the correct entries $\hat{\tilde{\alpha}}_{t,i}$ are consistently set to zero {(as $T$ goes to infinity)}. However, standard LASSO approaches do not necessarily exhibit these properties, as shown in \cite{zou2006adaptive}. The authors suggest a pre-estimation step for $\hat{\xvec{\alpha}}_i$, which can be done by an ordinary least squares (OLS) or ridge-regression.\footnote{The penalty parameter {in the ridge regression} can be chosen by a cross-validation study, e.g. as implemented in the R-package \texttt{glmnet}.} The estimated coefficients of this pre-estimation step are then employed as weighting factors $\xvec{\varpi}_i$ for {$\tilde{\xvec{\alpha}}_i$}. {The estimates of the pre-estimation step are denoted by $\xvec{\hat{\alpha}}^{(*)}_i$, which are obtained by ridge-regression in our case. Then,} the estimates are given by
\begin{equation}\label{eq:adalasso}
\hat{\tilde{\xvec{\alpha}}}_i \;\; \in \;\; \argmin_{\tilde{\xvec{\alpha}}_i} \left(||\xmat{K}\tilde{\xvec{\alpha}}_i - \xvec{y}_{\cdot i}||^2_2 + \lambda_a  ||\xvec{\varpi}_i \circ \tilde{\xvec{\alpha}}_i||_1 \right) \, ,
\end{equation}
where the observed process is $\xvec{y}_{\cdot i}$, the $L^p$-norm is denoted by $||\cdot||_p$, and $\circ$ is the Hadamard product. The elements of the weights vectors $\xvec{\varpi}_{i} = ({\varpi}_{t,i})_{t = 1, \ldots, T}$ are specified as ${\varpi}_{t,i} = \frac{1}{|\hat{\alpha}^{(*)}_{t,i}|^\gamma}$, where $\gamma > 0$ is a tuning parameter influencing the threshold function.\footnote{Through the parameter $\gamma$, the impact of larger estimated values $\hat{\xvec{\alpha}}^{(*)}_i$ on the penalty term can be strengthened or diminished. For use in our setting, we suggest setting $\gamma = 1$.} This convex optimization problem {in \eqref{eq:adalasso}} can be solved by corresponding optimization approaches, such as the LARS-algorithm of \cite{efron2004least}. Furthermore, let $\hat{\xvec{\alpha}}_i = \xmat{K}\hat{\tilde{\xvec{\alpha}}}_i$ be the estimated overall mean levels.
Given an appropriate choice of the penalization parameter $\lambda_a$,\footnote{To select $\lambda_a$, a $T$-fold cross-validation (CV) should be performed over a grid of possible values of $\lambda_a$. This approach is readily implemented, for instance, in the R-package \texttt{glmnet} (see \citealt{package_tibshi2010}).} the {estimator of} $\xvec{\alpha}_i$ does exhibit the oracle-properties, i.e., consistency in variable selection for each individual location, as well as asymptotic normality with an expectation of zero differences $\hat{\xvec{\alpha}}_i - \xvec{\alpha}_i$ as $T$ goes to infinity. This ensures that the true features are asymptotically selected. This is important because change points should not be overlooked in the first stage.

It is important to highlight at this point that $\hat{\xvec{\alpha}}_i$ consistently estimates the true, overall mean of location $i$ including any spatial spillover effect, even though the spatial dependence has been ignored. For each of these overall mean levels, we select all time points $t$, for which $\hat{\alpha}_{t,i} \neq \hat{\alpha}_{t-1,i}$ and consider them as candidate change points for $\xvec{a}_t$ in the second step. {This yields our set of candidate change points $\xcal{T}_i^{(1)} = \{\tau: \hat{\alpha}_{\tau,i} \neq \hat{\alpha}_{\tau-1,i} \}$.}

{It is worth noting that} a breakpoint in one location is {usually} reflected into other locations due to the spatial dependence. Thus, we tend to overestimate the true amount of changes in the individual series. {That is, there are more changes in $\xvec{\alpha}_t$ (i.e., blue curves in Figure \ref{fig:example}) than changes in the local mean levels $\xvec{a}_t$ (i.e., red curves in Figure \ref{fig:example})}. Overestimating - in contrast to underestimating - the candidate change points is less severe, as in the second step, the LASSO estimator can still eliminate incorrectly specified change points. In contrast, if a change is not detected in the first step, it will completely be eliminated from the model and cannot be detected in the second stage. {One might therefore ask whether ignoring spatial dependence can lead to missing structural breaks in the first step (e.g. because two opposite mean changes cancel each other out).} This can only happen if and only if $\hat{\alpha}_{t,i} = \hat{\alpha}_{t-1,i}$ for all locations $i = 1, \ldots, n$, but there exists at least one location $k$, for which $a_{t,k} \neq a_{t-1, k}$. However, {if there is a change in the local mean levels {$\xvec{a}_t$}, then this must be reflected into the overall mean levels {$\xvec{\alpha}_{t}$}.} This is shown in Proposition \ref{prop:first_stage}, for which the proof can be found in the Appendix. As a consequence, we obtain selection-consistent estimates of the candidate change points. 

\begin{proposition}\label{prop:first_stage}
If $\varrho(\xmat{W}) < 1$, all changes in $\xvec{a}_t$ are reflected into the overall mean level $\alpha_{t,i}$ of at least one location $i$.
\end{proposition}

{Moreover, when} applying the first stage estimation, we therefore recommend to analyze the sparsity level associated to the chosen $\lambda_a$ sequence. If the number of detected parameters, i.e., the number of potential change points, is too large then we suggest to use a sparser decision criterion, such as the sparsest solution which does not significantly differ from the minimum out-of-sample error and vice versa.

\subsection{Second stage: estimation of the full model}

In the second stage, we estimate the full model, i.e., both the changes in the mean {$\xvec{a}_{t}$} as well as the spatial dependence $\xmat{W}$. For this estimation, we propose again using a LASSO approach.
In particular, \eqref{eq:model} can be rewritten in a linear form as
\begin{equation}\label{reg_2nd}
\xvec{Y} = \xmat{\Psi} \tilde{\xvec{a}} + \xmat{Z} \xvec{\xi} + \xvec{\varepsilon} \, ,
\end{equation}
where {$\xvec{Y}=vec(\xmat{\Upsilon})$, $\xvec{\varepsilon}= (\xvec{\varepsilon}_1, \ldots, \xvec{\varepsilon}_T)'$} is the $nT$-dimensional vector of the stacked errors, $\xmat{Z} = \xmat{I}_n \otimes \xmat{\Upsilon}$ and $\xmat{\Psi}$ is a lower triangular, block diagonal matrix. The vectorization operator is denoted by {$vec$}, and $\otimes$ is the Kronecker product. To be precise, $\xmat{\Psi}$ is the direct sum of $n$ triangular matrices $\xmat{K}$ of dimension $T \times T$. {The vector of coefficients $\tilde{\xvec{a}}$ is defined as changes in $\xvec{a}$, i.e.,} {$\xvec{a} = \xmat{\Psi}\tilde{\xvec{a}} = (\xvec{a}_{1}, \ldots, \xvec{a}_{T})' = vec(\xmat{A})$ with} $\xmat{A} = (a_{t,i})_{t = 1, \ldots, T; i = 1, \ldots, n}$. 
In addition, we assume that $\tilde{a}_{t,i} = 0$ for {$t \notin \xcal{T}_i^{(1)}$} and all $i = 1, \ldots, n$. That said, the coefficients are forced to be zero for all time points for each location $\xvec{s}_i$, where no change is expected, which is based on the results of the first stage.

Aside from the mean-level term $\xmat{\Psi} \tilde{\xvec{a}}$, we account for possible spatial dependence by $\xmat{Z} \xvec{\xi}$. The coefficients $\xvec{\xi}$ are the vectorized spatial weights, i.e.,  $\xvec{\xi} = vec(\xmat{W})$. It is worth noting that the diagonal elements of $\xmat{W}$ are assumed to be zero, thus allowing us to avoid self-dependencies, and the number of estimated parameters in $\xmat{Z} \xvec{\xi}$ is reduced.

Eventually, the estimated coefficients of the full model are given by the optimization problem
\begin{footnotesize}
\begin{align*}\label{adalasso2}
(\hat{\tilde{\xvec{a}}}, \hat{\xvec{\xi}})' \;\; \in \;\; \argmin_{(\tilde{\xvec{a}}, \xvec{\xi})'} \; &
||\xmat{Z} \xvec{\xi} + \xmat{\Psi}\tilde{\xvec{a}} - \xvec{y}||^2_2 + \lambda_b \, || \xvec{\xi}||_1 \, , \, \text{subject to:} \\
                        & \tilde{a}_{t,i} = 0 \; \text{for all} \;  t \notin \xcal{T}_i^{(1)}, \, 1 \leq i \leq n; & \\
                        & w_{ij} \geq 0 \, \text{and} \, w_{ii} = 0 \; \text{for all} \; 1 \leq i,j, \leq n ; \\
                        & \sum_{j=1}^{n} w_{ij} \leq 1  \text{for all $i = 1, \ldots, n$}\, .
\end{align*}
\end{footnotesize}
The observed vector $\xvec{Y}$ is denoted by $\xvec{y}$. For this second stage, a constrained LASSO approach has been implemented, which ensures that  $\sum_{j=1}^{n} w_{ij} \leq 1$ for all $i$ and, therefore, the non-singularity of $(\xmat{I} - \xmat{W})$. This constraint mostly influences the estimation in cases of strong spatial dependence, when the spatial process is close to non-stationarity (i.e., $\varrho(\xmat{W})$ is close to one). Other constrained LASSO estimators have been proposed, for instance, by \cite{James12,James18}. {Moreover, the weights must be non-negative. Hence,} it is not feasible to model negative spatial dependence using our model, an occurrence that rarely arises in practice \citep[cf.][]{Cressie93}. Least squares estimators under the presence of spatial dependence would not generally be sign-consistent \citep[see][]{Lam13}. Thus, this restriction on positive spatial dependence {ensures} sign-consistency.

It is important to note that $(\hat{\tilde{\xvec{a}}}, \hat{\xvec{\xi}})'$ lies in a minimizing set, because the coefficients are not necessarily unique estimates, since the problem is high-dimensional. {In general}, it is challenging to find the true direction of the spatial dependence between two locations {in the high-dimensional setting}. {In our simulations, for these cases, we observed weaker, spurious entries $\hat{w}_{ij}$ on the mirrored side when the true non-zero weight is one-directional, i.e., $w_{ij} = 0$ and $w_{ji} > 0$.} Thus, we focus on the performance of our approach for {both symmetric and} non-symmetric weighting schemes in a series of simulation studies in Section \ref{sec:MC}. {We refer to symmetry in the positive weights. That is, $\xmat{W}$ is called symmetric if all entries are two-sided but these pairs of links are not necessarily identical, and non-symmetric if there are directional/one-sided links.} If $\xmat{W}$ is assumed to be symmetric, the number of estimated parameters would reduce to $\frac{1}{2}n(n-1) + nT$ (cf. \citealt{Bhattacharjee13}).

 As there is a strong interdependence between $\xvec{a}_t$ and $\xmat{W}$ (i.e., if $\xvec{a}_t$ is estimated correctly, then also the spatial dependence structure will be estimated correctly, and vice versa), it {seems to be} sufficient to only penalize the estimated weights of $\xmat{W}$, but not the local mean levels $\xvec{a}_t$. {In all our simulations, we found that there was no need for two penalty parameters because the one included controlled for both the sparseness in the weights and implicitly in the local means. Already in the first step the mean changes were penalized, such that the potential change points are restricted to ones found in the first stage.} Thus, only one penalty parameter $\lambda_b$ appears in the objective function. Nonetheless, adding a second penalty $\lambda_{a2}$ for $\xvec{a}_t$ is possible and can lead to a higher sparsity of $\hat{\xvec{a}}_t$ at the cost of computational effort. The second-stage optimization could be computationally implemented using, for instance, the \texttt{ROI} solver with the \texttt{qpoases}-solver plugin (see \cite{ROI} for details).\footnote{All code and data to reproduce our results are provided with this manuscript.}

\subsection{Selection of the penalty parameter $\lambda_b$}

Of course, the selection of the penalty parameter $\lambda_b$ is crucial for the goodness-of-fit. In Figure \ref{fig:lambda_selection}, the mean absolute {error} of the estimated matrix $\hat{\xmat{W}}$ and the estimated local mean levels $\hat{\xvec{a}}_t$ are plotted by red and blue circles, respectively. The penalty parameter $\lambda_b$ discretely varies over an exponentially decaying grid and the OLS-solution has been added at $50$ ($\lambda_b = 0$).

For a series of simulations, which are explained in the next section in more details, we observe that the true model with {minimal errors} is on the sequence of considered values of $\lambda_b$, the so-called $\lambda_b$-path.  Moreover, the {best} estimates for $\xmat{W}$ and $\xvec{a}_t$ are obtained for the same $\lambda_b$ {amongst those considered} in almost all cases. That is, if the structure of $\xmat{W}$ is estimated correctly, so $\xvec{a}_t$ is estimated correctly and vice versa. {Thus, the minima of the blue and red curves} are very close or coincide with each other {in Figure \ref{fig:lambda_selection}}.

The best parameter $\lambda_b$ should be selected by a certain goodness-of-fit criterion in a cross-validation study {(CV)}. However, in our case, common methods, like minimizing mean absolute errors (MAE) or root mean squared errors (RMSE) of the predicted values $\hat{Y}$ does not lead to the best value of $\lambda_b${, because we would not account for the spatial autocorrelation in the errors.} The RMSE {of a random $k$-fold CV} and its minimal value are depicted as gray curves in Figure \ref{fig:lambda_selection}. Apparently, this classical approach does not lead to a good selection in our spatiotemporal setting\footnote{All details of our simulation setup are explained in Section \ref{sec:MC}.}.

Thus, we propose a completely novel performance measure, which exploits the fact that the spatial dependence is assumed to be constant over time. More precisely, we compare the sample spatial autocorrelation and the spatial autocorrelation estimated by the model. {In this way, we incorporate the information of the spatial autocorrelation (i.e., $(\xmat{I} - \xmat{W})^{-1} \xvec{\varepsilon}_t$ in \eqref{eq:model2}) to select the best-fitting spatial weight matrix. In turn, this enables the estimation of the local mean levels $\xvec{a}_t$.} {Thus, as mentioned before,} it is sufficient if the selection criterion is only based on the spatial dependence, as the minimal {error} of $\hat{\xmat{W}}$ occurs at the same $\lambda_b$ position as the minimal {error} of $\hat{\xvec{a}}_t$. 

For the cross-validation study, we first estimated the model using all odd time points and validated the model using all even time points. That implicitly assumes that change points could only occur every two time steps. Then, the model validation is repeated for all odd time points. The sample spatial autocorrelation can be obtained from the sample covariance matrix from the mean-adjusted observations, that is,
\begin{equation*}
  \xvec{Z}_t^{s} = \xvec{Y}_{t\cdot} - (\xmat{I} - \hat{\xmat{W}})^{-1} \hat{\xvec{a}}_t \, .
\end{equation*}
Further, the estimated spatial autocorrelation {of the model} can easily be obtained from $\hat{\xmat{W}}$, e.g., via simulation of new random errors $\xvec{\varepsilon}_t^{*}$ from the distribution $f_\varepsilon$. Alternatively, the model's spatial autocorrelation $(\xmat{I} - \hat{\xmat{W}})^{-1} \xmat{\Sigma}_\varepsilon (\xmat{I} - \hat{\xmat{W}}')^{-1}$ with the error covariance matrix $\xmat{\Sigma}_\varepsilon$ could directly be computed or parametric bootstrap procedures could be applied (cf. \citealt{Efron85,Efron97}). For our setting, we {re-simulated} the errors {$\xvec{\varepsilon}_t^{*}$} independently {from} the same normal distribution with a homoscedastic variance. {Then, we get}
\begin{equation*}
  \xvec{Z}_t^{m} = (\xmat{I} - \hat{\xmat{W}})^{-1} \xvec{\varepsilon}_t^{*} .
\end{equation*}

 Since we only compare the spatial autocorrelation, the variance of {$\xvec{\varepsilon}_t$ (and $\xvec{\varepsilon}_t^{*}$)} does not affect the selection criterion
\begin{equation*}
  \text{MAE}(Corr(\xvec{Z}_t^{s}), Corr(\xvec{Z}_t^{m})) \, .
\end{equation*}
In Figure \ref{fig:lambda_selection}, this new criterion is depicted by the {orange} line, which numerically has its minimum close to the best model (deviation from the best $\lambda_b$ is less than 1 value in the exponentially decaying grid).

\begin{figure}
  \centering
  \input{Selection_lambda_rho.tex}
  \caption{Selection of the penalty parameter $\lambda_b$. The MAE of the estimates $\hat{\xmat{W}}$ and $\hat{\xvec{a}}_t$ are represented by the dotted curves with empty circles (right axis, blue) and triangles (left axis, red), respectively. Moreover, the out-of-sample RMSE of the predicted values $\hat{\xvec{Y}}$ obtained in a {random $k$-fold} CV study is drawn as dashed curve (black) and the newly introduced {selection criterion} is drawn as solid line (orange). The values of $\lambda_b$, which would be selected by these two criteria (i.e., the minima of the dashed or solid curves) are depicted by the dashed/solid vertical lines. The columns refer to the chosen weighting scheme $\tilde{\xmat{W}}_1$, $\tilde{\xmat{W}}_2$, or $\tilde{\xmat{W}}_3$, while the rows refer to the strength of the spatial dependence from weak, moderate, to large.}\label{fig:lambda_selection}
\end{figure}

\section{Monte Carlo Simulations}\label{sec:MC}

In this section, we evaluate the performance of the LASSO estimators by a series of Monte Carlo simulations. First, we analyze (a) how well the algorithm detects the existence of spatially dependent observations, (b) how well the algorithm detects spatially independent observations, and (c) how well the changes in the mean level are captured by the approach. We perform these simulations for different specifications of the true underlying spatial dependence $\xmat{W}$. Secondly, we analyze how the performance between different levels of spatial dependence differs. 

Generally, one would expect the LASSO estimators to perform more poorly for small spatial dependence, as the resulting effects are weaker. Moreover, if $\varrho(\xmat{W})$ is close to 1, we could be faced with difficulties as the maximum of the objective function would be close to the constraint (i.e., close to regions where the model is not well-defined). Thus, we evaluate the performance for weak ($||\xmat{W}||_{\infty} = 0.25$), moderate ($||\xmat{W}||_{\infty} = 0.5$), and large spatial dependence ($||\xmat{W}||_{\infty} = 0.75$).

\begin{figure}
  \centering
  \begin{tabular}{c c c}
\hline
$\tilde{\xmat{W}}_1$ & $\tilde{\xmat{W}}_2$ & $\tilde{\xmat{W}}_3$ \\
\hline
\multicolumn{3}{c}{\emph{True weight matrix}} \\
  {\includegraphics[width = 0.25\textwidth]{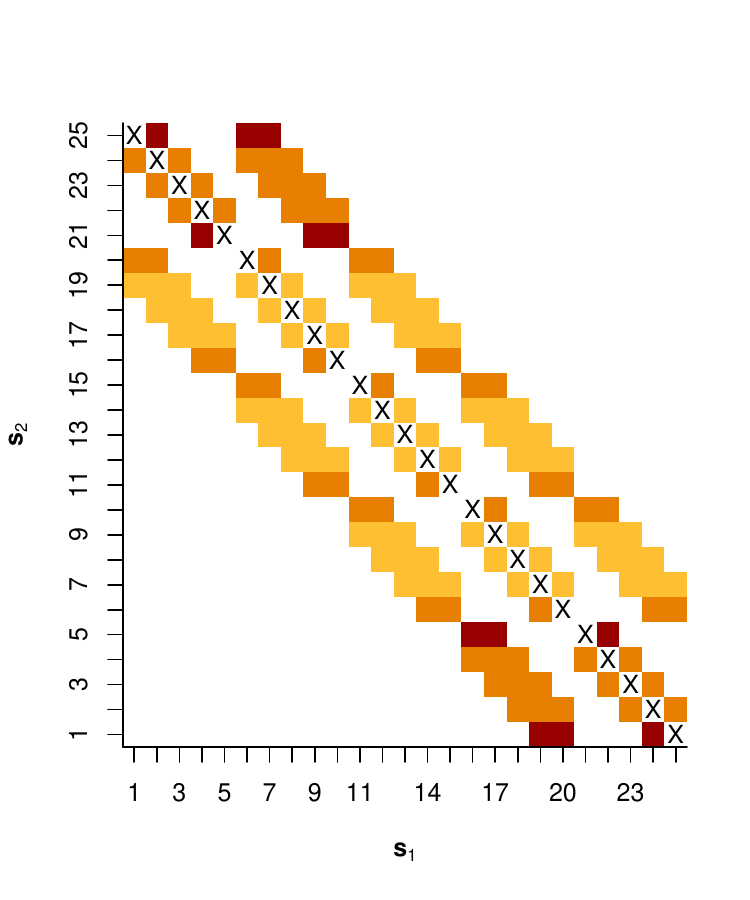}} &
  {\includegraphics[width = 0.25\textwidth]{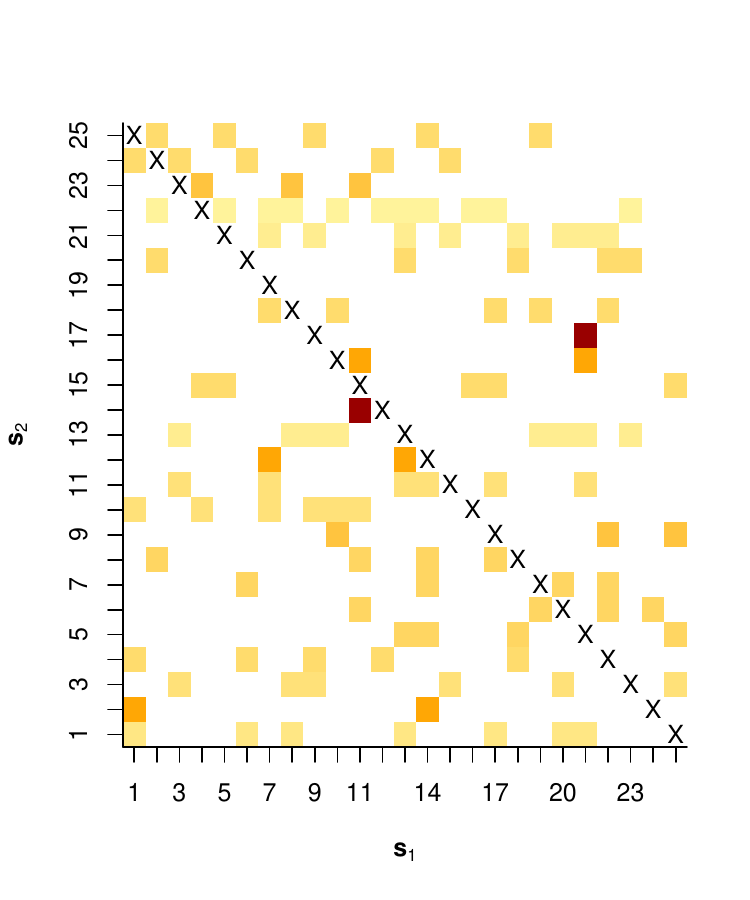}} &
  {\includegraphics[width = 0.25\textwidth]{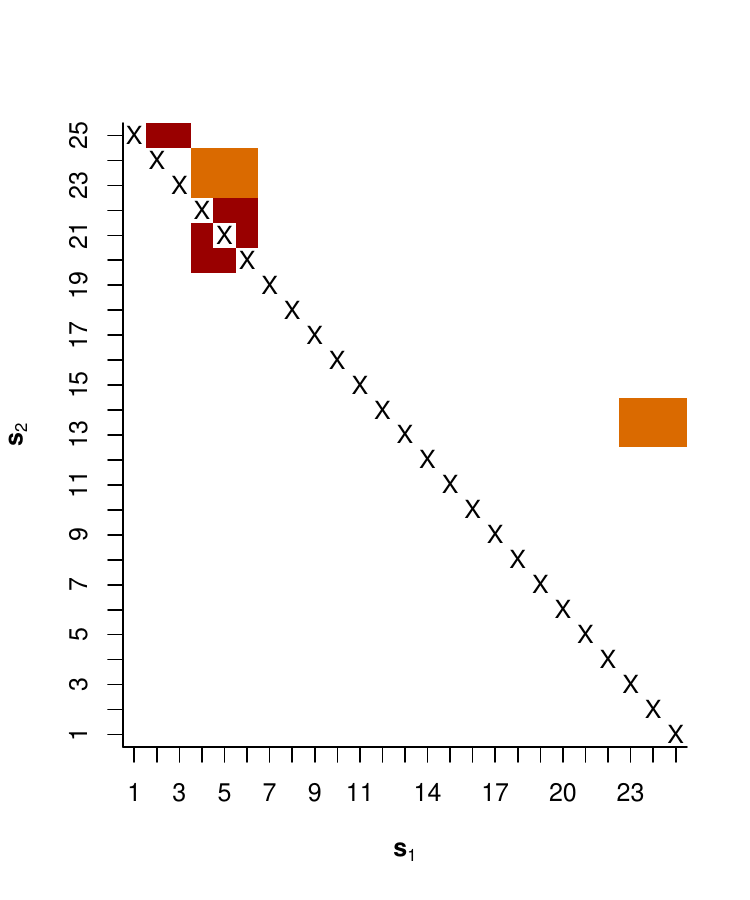}} \\
\multicolumn{3}{c}{\emph{Estimated weight matrix}} \\
  {\includegraphics[width = 0.25\textwidth]{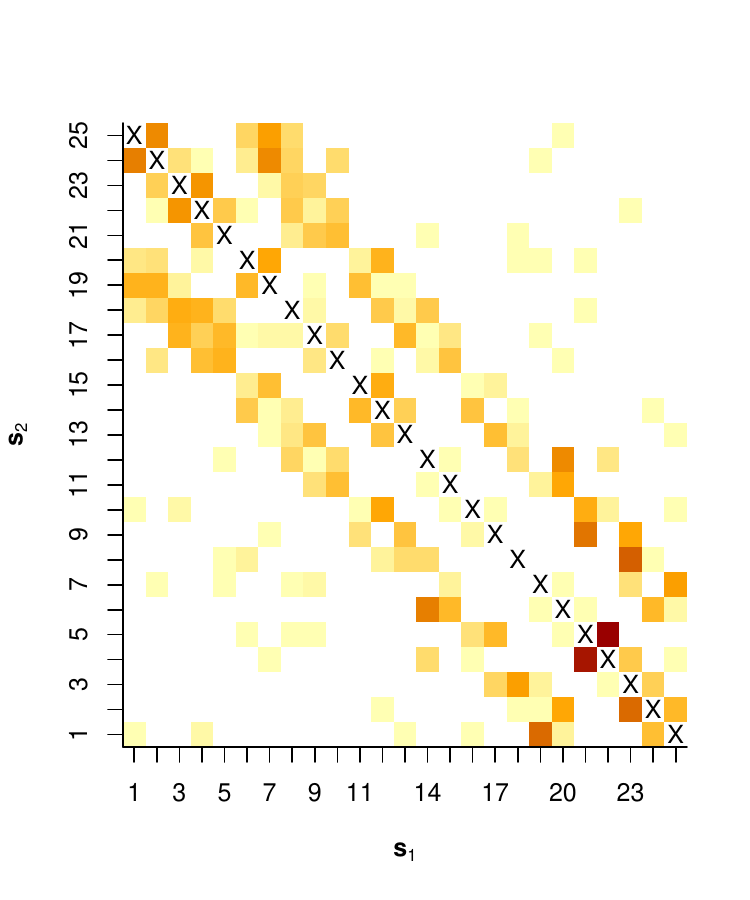}} &
  {\includegraphics[width = 0.25\textwidth]{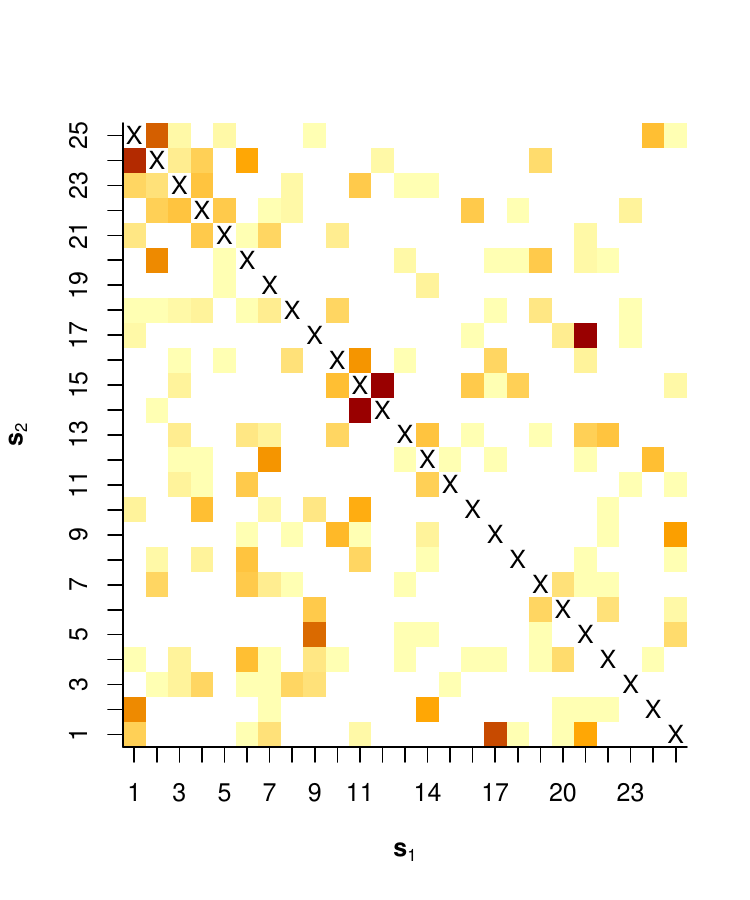}} &
  {\includegraphics[width = 0.25\textwidth]{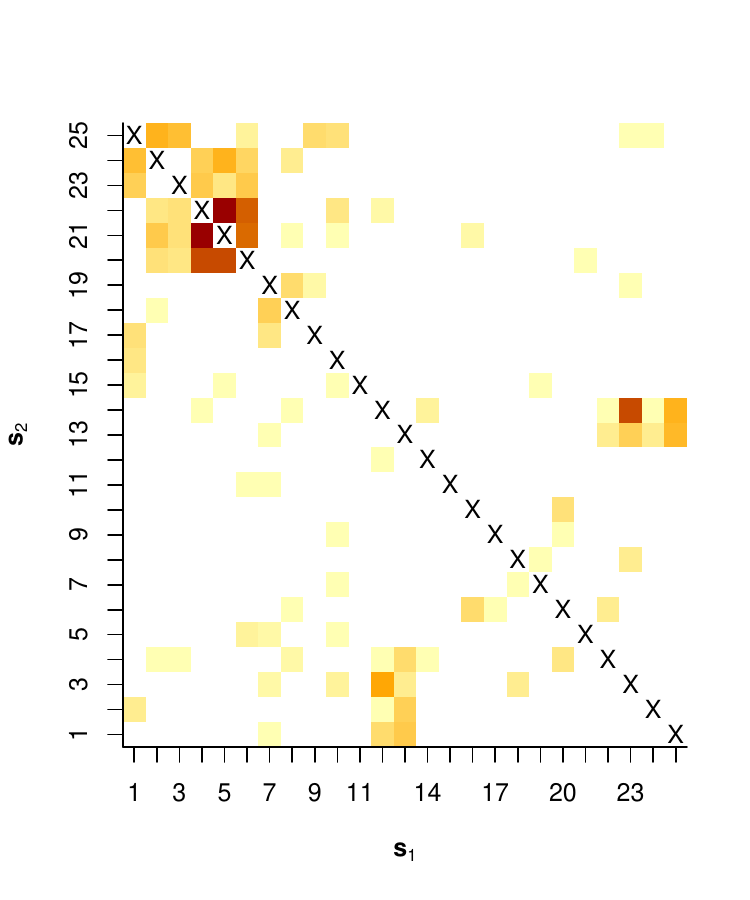}} \\
\hline
  \end{tabular}
  \caption{First row: True spatial weight matrices $\tilde{\xmat{W}}$ considered in the simulation study, i.e., a row-standardized Queen's contiguity matrix (left, $\tilde{\xmat{W}}_1$), row-standardized randomly sampled spatial weights (center, $\tilde{\xmat{W}}_2$), and row-standardized block structure (right, $\tilde{\xmat{W}}_3$). In the second row, we plot the estimated counterparts $\hat{\xmat{W}}$, following the principle that the darker the color, the higher the true/estimated value of the link. White entries represent a link that is zero or is estimated to be exactly zero, respectively.}\label{fig:W_ex}
\end{figure}

The spatiotemporal process is simulated on a spatial lattice $D = \{ \xvec{s} \in \xset{Z}^2 : 0 < s_1 \leq 5, 0 < s_2 \leq 5 \}$, i.e., $n = 25$, for discrete time points $t = 1, \ldots, T$ with $T \in \{100, 200\}$. To define the spatial dependence, we consider three row-standardized weighting schemes $\tilde{\xmat{W}}$, as illustrated in the first row of Figure \ref{fig:W_ex}. First, we consider a Queen's contiguity matrix $\tilde{\xmat{W}}_1$, which is depicted on the left-hand side. For this scheme, dependencies cannot be found in areas that are far away. It is also assumed that if there is a connection from area $i$ to $j$, then there must be a feedback from $j$ to $i$ as well. This specific structure is quite commonly used in empirical research; for instance, used in \cite{Buettner99,Ross07,Tsai09}. Secondly, a randomly sampled spatial dependence structure $\tilde{\xmat{W}}_2$ is considered, in which a link between two locations exists with a probability of 20 percent. Thus, there would not necessarily be a connection from $j$ to $i$ if $i$ and $j$ are connected. Note that this matrix is also row-standardized, such that $||\tilde{\xmat{W}}_2||_{\infty} = 1$. Thirdly, we assume that only a few locations are dependent in $\tilde{\xmat{W}}_3$. Moreover, these weights should appear in close proximity. Thus, we randomly assigned 3 blocks of minimal size $1 \times 1$ up to maximal size $5 \times 5$ with positive weights. These blocks are not necessarily quadratic, because we independently sampled both lengths from the set $\{1, ..., 5\}$ as well as the offset point in the top left corner. In contrast to the contiguity matrix, the dependent observations can but must not occur in the closest neighborhood depending on the indexing of the locations. All spatial weighting matrices $\tilde{\xmat{W}}$, which have random features, are sampled again for every iteration of the simulation. Finally, $\xmat{W} = \rho \tilde{\xmat{W}}$ with $\rho \in \{0.25, 0.5, 0.75\}$. {Thus, in total, we considered $3 \times 3 \times 2 = 18$ different settings, for which the spatial dependence randomly changes in each replication for $\tilde{\xmat{W}}_2$ and $\tilde{\xmat{W}}_3$.} {For each of the 18 settings, we} simulate the process {with} $M = 2^{9} = 512$ replications.

{There is at least one change point in each location.} To create realistic conditions, {the locations are divided} into two groups, so that the change points do not occur at the same {time points across space}. This results in a first group containing 10 locations having two changes in the mean at $\tau_1 = 0.5 T$ and $\tau_2 = 0.75 T$, whereas only one structural break occurs at $\tau_3 = 0.25 T$ for the second group, i.e., the remaining 15 locations. Consequently, there are three mean changes at equidistant time points. {For} the locations of the first group, the zero-mean level increases to $3$ at $\tau_1$ and decreases again to $0$ at $\tau_2$. In contrast, the mean level is $0$ for $t < \tau_3$ and $7$ for $t \geq \tau_2$ for the locations in the second group. The residuals $\xvec{\varepsilon}_t$ are independently sampled normally distributed random variables with zero mean and unit variance.  

{To evaluate the performance, we consider the following criteria.} First, we compute the average bias of the non-diagonal {spatial weights}
\begin{equation*}
\mathrm{B}_w = \frac{1}{n^2-n}\sum_{\substack{i,j = 1 \\ i\ne j}}^n  (\hat{w}_{ij} - w_{ij})  \,
\end{equation*}
to assess the estimation accuracy of the weights $\hat{w}_{ij}$ {in terms of the average deviation from their true values}. It is important that the diagonal entries are excluded in order to avoid any unjustified improvement of our results. Furthermore, the MAE of the {estimated weights}
\begin{equation*}
\mathrm{MAE}_w = \frac{1}{n^2-n}\sum_{\substack{i,j = 1 \\ i\ne j}}^n  | \hat{w}_{ij} - w_{ij} | \,
\end{equation*}
are computed as measure of estimation precision.

{Second}, we report the specificity and sensitivity of all elements of $\xmat{W}$ neglecting {again} the diagonal elements. The specificity provides information regarding the amount of correctly {excluded} links, i.e., the proportion of correctly estimated zero entries in $\hat{\xmat{W}}$, that is,
\begin{equation*}
\Pi_0 = \frac{ | \{ (i,j)' : w_{ij} = 0 \wedge \hat{w}_{ij} = 0 ,i \neq j \} | }{ | \{ (i,j)' : w_{ij} = 0 , i \neq j\} |},
\end{equation*}
or true negative rate. Indeed, using this criterion alone cannot yield meaningful results, as na\"{i}ve methods, such as setting all weights to zero, leads to a specificity of 1, even though such a strategy would create numerous false identifications for links that were actually not zero. Hence, the sensitivity
\begin{equation*}
\Pi_w = \frac{  | \{ (i,j)' :  w_{ij} > 0 \wedge \hat{w}_{ij} > 0 , i \neq j\} | }{ | \{ (i,j)' : w_{ij} > 0 , i \neq j\} |}
\end{equation*}
should also be considered to assess the performance for correctly {selected} positive weights (i.e., true positive rate).

{Third}, we examine the average bias and {MAE} for the estimated mean levels $\xvec{a}_t$. The average bias is defined as
\begin{equation*}
\mathrm{B}_a = \frac{1}{nT} \sum_{i=1}^{n} \sum_{{t=1}}^{T} (\hat{a}_{t,i} - a_{t,i})
\end{equation*}
along with the MAE
\begin{equation*}
\mathrm{MAE}_a = \frac{1}{nT} \sum_{i=1}^{n} \sum_{{t=1}}^{T} | \hat{a}_{t,i} - a_{t,i} | \, .
\end{equation*}

{Fourth, we consider the root mean square error (RMSE) of the fitted process, that is
\begin{equation*}
\mathrm{RMSE}_y = \sqrt{\frac{1}{nT} \sum_{i=1}^{n} \sum_{{t=1}}^{T} ( \hat{y}_{t,i} - y_{t,i} )^2} \, .
\end{equation*}
}

\begin{table}
\centering
\caption{Results of the simulation study with $M=2^9$ simulations, $n = 25$, and $T \in \{100,200\}$. We report the results for different magnitudes of spatial dependence $\rho \in \{0.25, 0.5, 0.75\}$, where $\xmat{W} = \rho \tilde{\xmat{W}}$.} \label{tab:sim_results}
\begin{footnotesize}
\begin{tabular}{l  l  c  c  c  c  c  c }
\hline
          & & \multicolumn{3}{c}{$T = 100$} & \multicolumn{3}{c}{$T = 200$} \\
          & & $\tilde{\xmat{W}}_1$ & $\tilde{\xmat{W}}_2$ & $\tilde{\xmat{W}}_3$ &$\tilde{\xmat{W}}_1$ & $\tilde{\xmat{W}}_2$ & $\tilde{\xmat{W}}_3$  \\
\hline
\multicolumn{8}{c}{\emph{weak spatial dependence} ($\rho = 0.25$)} \\
\parbox[t]{2mm}{\multirow{4}{*}{\rotatebox[origin=c]{90}{Weights $\hat{\xmat{W}}$}}}
                & $\mathrm{B}_w$       &  0.000    &  0.000     &  0.003     &  0.000     &  0.000     &  0.002       \\
                & $\mathrm{MAE}_w$     &  0.013    &  0.016     &  0.009     &  0.011     &  0.014     &  0.008       \\
                & $\Pi_0$ (spec.)      &  0.818    &  0.759     &  0.817     &  0.804     &  0.728     &  0.799       \\
                & $\Pi_w$ (sens.)      &  0.478    &  0.379     &  0.401     &  0.623     &  0.485     &  0.533       \\[.2cm]
\parbox[t]{2mm}{\multirow{2}{*}{\rotatebox[origin=c]{90}{Mean $\hat{\xvec{a}}_t$}}}
                & $\mathrm{B}_a$       & -0.023    &  0.008     &  0.241     & -0.039     & -0.062     &  0.173       \\
                & $\mathrm{MAE}_a$     &  0.703    &  0.765     &  0.704     &  0.525     &  0.597     &  0.548       \\[.2cm]
                & $\mathrm{RMSE}_y$    &  0.975    &  0.975     &  0.982     &  0.983     &  0.982     &  0.988       \\[.3cm]
\multicolumn{8}{c}{\emph{moderate spatial dependence} ($\rho = 0.5$)} \\
\parbox[t]{2mm}{\multirow{4}{*}{\rotatebox[origin=c]{90}{Weights $\hat{\xmat{W}}$}}}
                & $\mathrm{B}_w$       &  0.000    &  0.001     &  0.002     &  0.000     &  0.001     &  0.001       \\
                & $\mathrm{MAE}_w$     &  0.019    &  0.028     &  0.013     &  0.015     &  0.024     &  0.010       \\
                & $\Pi_0$ (spec.)      &  0.812    &  0.660     &  0.797     &  0.834     &  0.641     &  0.792       \\
                & $\Pi_w$ (sens.)      &  0.744    &  0.611     &  0.593     &  0.865     &  0.736     &  0.719       \\[.2cm]
\parbox[t]{2mm}{\multirow{2}{*}{\rotatebox[origin=c]{90}{Mean $\hat{\xvec{a}}_t$}}}
                & $\mathrm{B}_a$       & -0.010    &  0.295     &  0.251     & -0.012     &  0.197     &  0.170       \\
                & $\mathrm{MAE}_a$     &  1.170    &  1.320     &  0.972     &  0.874     &  1.072     &  0.782       \\[.2cm]
                & $\mathrm{RMSE}_y$    &  0.968    &  0.952     &  0.975     &  0.970     &  0.959     &  0.981       \\[.3cm]
\multicolumn{8}{c}{\emph{large spatial dependence} ($\rho = 0.75$)} \\
\parbox[t]{2mm}{\multirow{4}{*}{\rotatebox[origin=c]{90}{Weights $\hat{\xmat{W}}$}}}
                & $\mathrm{B}_w$       &  0.000    &  0.001     &  0.001     &  0.000     &  0.000     &  0.000       \\
                & $\mathrm{MAE}_w$     &  0.022    &  0.037     &  0.015     &  0.017     &  0.032     &  0.012       \\
                & $\Pi_0$ (spec.)      &  0.797    &  0.595     &  0.801     &  0.878     &  0.627     &  0.809       \\
                & $\Pi_w$ (sens.)      &  0.875    &  0.756     &  0.690     &  0.949     &  0.847     &  0.793       \\[.2cm]
\parbox[t]{2mm}{\multirow{2}{*}{\rotatebox[origin=c]{90}{Mean $\hat{\xvec{a}}_t$}}}
                & $\mathrm{B}_a$       &  0.071    &  0.294     &  0.195     & -0.002     & -0.057     &  0.110       \\
                & $\mathrm{MAE}_a$     &  1.840    &  1.876     &  1.199     &  1.307     &  1.524     &  0.981       \\[.2cm]
                & $\mathrm{RMSE}_y$    &  0.958    &  0.940     &  0.970     &  0.953     &  0.941     &  0.977       \\
\hline
\end{tabular}
\end{footnotesize}
\end{table}

All results obtained are reported in Table \ref{tab:sim_results}. First, it is possible to estimate the underlying spatial dependence for spatiotemporal autoregressive models, even with the presence of an undefined number of mean changes. This is a substantial advantage compared to classical approaches where {structure of the spatial dependence, i.e., the spatial weight matrix, is pre-specified by the user.} Generally, the proposed penalized regression method shows a good performance for estimating this matrix $\xmat{W}${, even in the case of asymmetric dependence structures, which are difficult to find with the classical approach as the true asymmetric weight matrix must be included in the set of candidates.} However, the performance is slightly dependent on the type of the true underlying spatial dependence structure. For matrices like the Queen's contiguity matrix where we have symmetric links, the algorithm can easily detect the relevant links. On average, 86.5 percent of all true links were detected for $\rho = 0.5$ and $T = 200$. For matrices $\tilde{\xmat{W}}$, which do not have symmetric entries, the performance is more difficult to evaluate. In particular, the algorithm estimates an undirected dependence, i.e., symmetric entries, rather than the true directed or one-way dependence. This is due to the identifiability of the weight parameters in the case of only three mean changes. Thus, specificity and sensitivity are lower for $\tilde{\xmat{W}}_2$ and $\tilde{\xmat{W}}_3$ compared to the symmetric case (i.e., $\tilde{\xmat{W}}_1$).

This can also be seen in Figure \ref{fig:lambda_selection}, where the absolute {deviation} of $\hat{\xmat{W}}$ {from its true value} is slightly higher in the asymmetric cases (second and third column). The tendency to estimate symmetric spatial dependence can also be seen for the example matrices of one replication, as shown in Figure \ref{fig:W_ex}. {The} estimates $\hat{\xmat{W}}$  are depicted in the second row. Even though the algorithm detects the block structure in $\tilde{\xmat{W}}_3$, it also erroneously estimates its counterpart on the mirrored side of the main diagonal, because it is difficult to distinguish between one- and two-sided dependencies. Nevertheless, the weight of the wrongly classified link is usually weaker than the true one. When interpreting single weights in practice, one should bear this issue in mind. That is, even though a link has been estimated in both directions, the weaker one could be spurious.

In general, the estimation precision of the weights is increasing with increasing time horizon $T$, while the average bias is decreasing. That means that the estimation of the spatial dependence structure in terms of the magnitude of the weights $w_{ij}$ and the number of correctly selected positive weights (i.e., the active set) was getting more precise in all considered cases with an increasing time horizon. As we observed for the spatial dependence, the estimation precision of $\xvec{a}_t$ is higher in all cases for $T = 200$ compared to $T = 100$ (i.e., the MAE is decreasing with increasing $T$). In contrast, the average bias is increasing in some of the cases, mostly if the spatial dependence is weak. In the majority of cases, the average bias of $\xvec{a}_t$ is, however, decreasing with an increasing number of time points. Generally, if we observe a deviation of the estimated levels $\hat{\xvec{a}}_t$ from the true ones, the difference is usually captured by the spatial dependence $\hat{\xmat{W}}$. Thus, local change points are erroneously reflected into other locations in some cases. That is, if the dependence is more precisely estimated, the estimated spatial mean level will be much closer to the true level $\xvec{a}_t$. In summary, it can be stated that the proposed estimation procedure can capture the dynamics of $\xvec{a}_t$ well, but it of course depends on the precision of the estimation of $\xmat{W}$.

\section{Real Data Example}\label{sec:empirical}

Housing prices generally depend on the location of the property and the surrounding housing prices. In more attractive regions that provide a better infrastructure, the housing prices are higher than in less-developed regions. In larger cities, the spatial dependence between housing prices of different districts or quarters is not always only due to spatial proximity, but also due to access to public transport connections, similar lifestyles, or cultural offerings. All these effects are challenging to measure or anticipate, as would be necessary for classical spatial econometric models. Moreover, we frequently observe that price levels fluctuate over time. These fluctuations could also be due to aspects of the legal framework, such as changes with property taxes or real-estate transfer taxes.

In the last section, we focus on the spatial interactions in regional housing prices. In particular, we analyze the condominium prices in the Berlin post code regions from January 1995 to December 2014. The data is taken from the Berlin committee of valuation experts for real estate (\emph{Gutachterausschuss}), which has information regarding the prices for each real estate transaction. To create a panel data set with equidistant time points, we pooled all contracts for a specific area within a month, observing 240 monthly prices over a period of 20 years. The prices are given in Euro per $m^2$, accounting for the currency changeover from Deutsche Mark to Euro in 2002. Regarding the spatial resolution, we define the locations by the respective {postal} codes, with only the first three digits being used. In total, there are 24 different unique 3-digit {postal codes} for the city of Berlin, which were constant over the entire time period studied. Each {post code} corresponds to exactly one unique self-contained area, except for {code} 140xx, which consists of two distinct spatial polygons. As there were more than 120 months with no valid {real-estate transaction} for the {spatial unit} 136xx, we excluded this area from our analysis. To gain numerical stability, the remaining 23 time series are transformed by a normal probability integral transformation {(PIT)}, {like in} \cite{uniejewski2018variance} who examined electricity prices{, to get closer to Gaussianity for a better stability of} our two-stage method. {Thus, the time series do} not exhibit extreme price spikes due to sparse transaction data for several months and/or areas. After the calculation of the relevant features, it {remains} possible to re-transform the series back into the original price domain.

To estimate $\xmat{W}$ and the mean changes $\xvec{a}_t$, we use the proposed method described in Section \ref{sec:LASSO}. We add one further assumption, namely that no change in $\xvec{a}_t$ occurred within the last 5 percent of the time points (i.e., all observations in 2014). Otherwise, the algorithm would have only one time point to estimate $\hat{\xvec{a}}_T$, two time points for $\hat{\xvec{a}}_{T-1}$, and so on. Hence, this restriction guarantees that there are at least 13 time points with which to estimate each local mean level $\xvec{a}_t$. This allows only for estimating a constant mean price level in the last year. Note that this local mean level may, however, vary across locations.\footnote{In general, we suggest to exclude at least five to ten time points at the end to get numerical stability of these estimated mean segments. However, note that it is not obligatory and could be neglected if one could expect changes in the last time points.}

\begin{figure}
  \centering
  \input{CP_Berlin_toprow.tex} \\
  \includegraphics[width=0.135\textwidth]{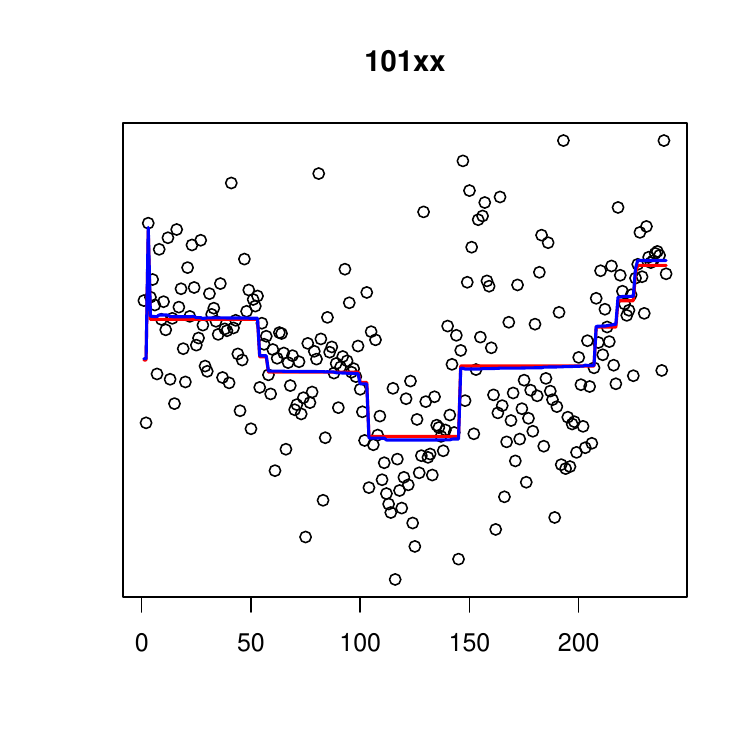}
  \includegraphics[width=0.135\textwidth]{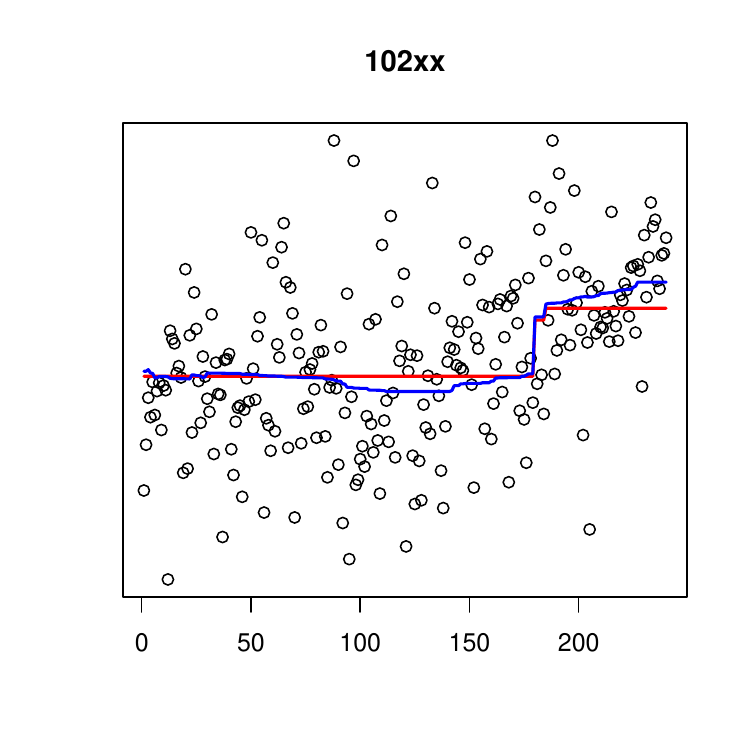}
  \includegraphics[width=0.135\textwidth]{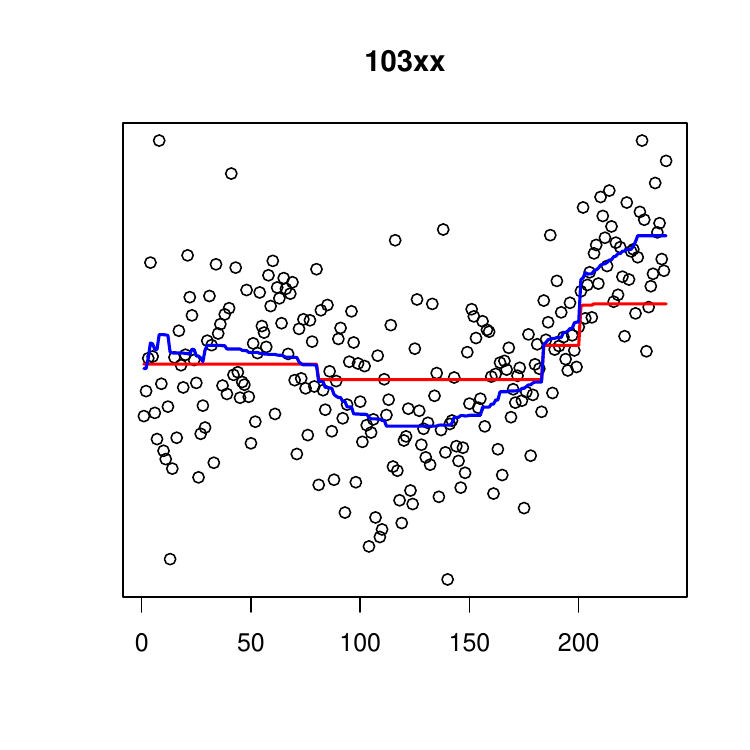}
  \includegraphics[width=0.135\textwidth]{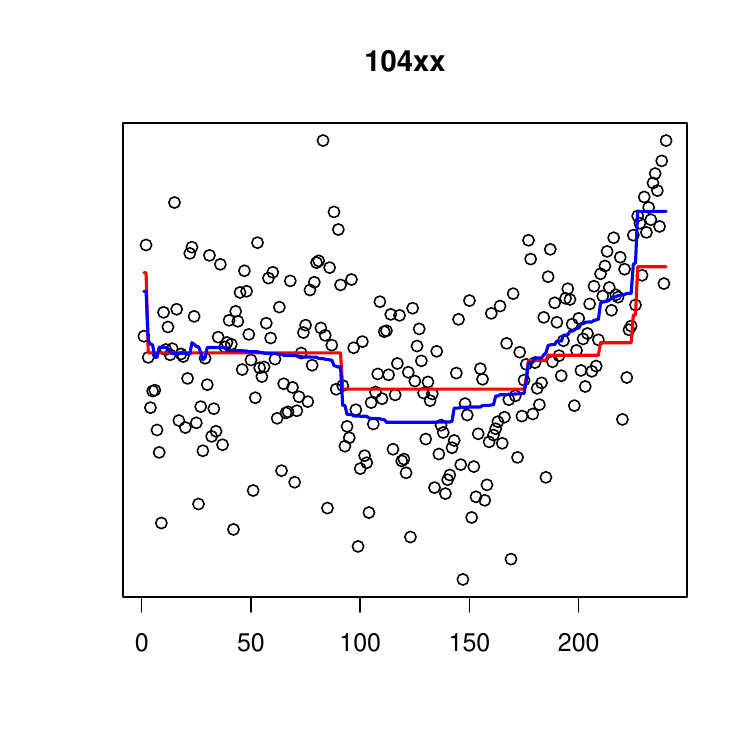}
  \includegraphics[width=0.135\textwidth]{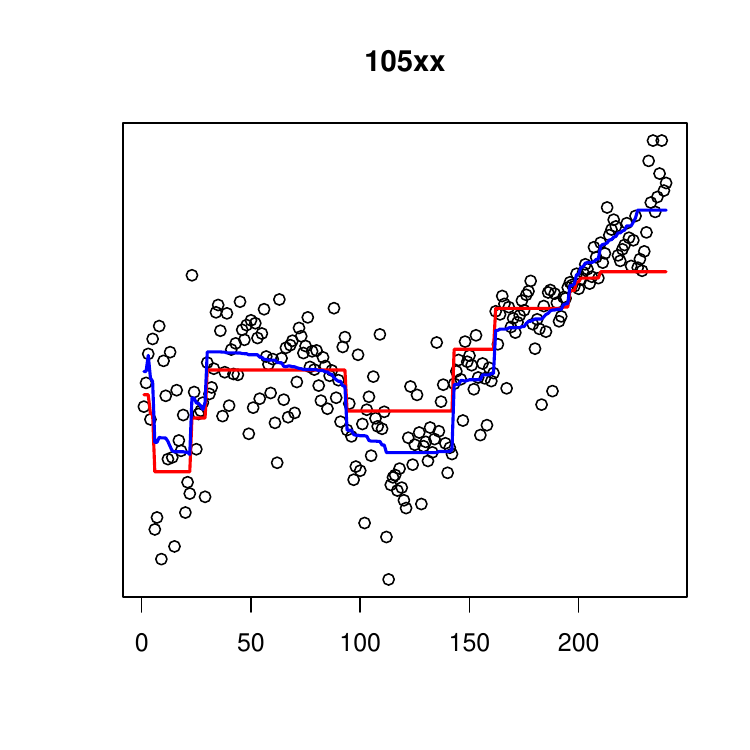}
  \includegraphics[width=0.135\textwidth]{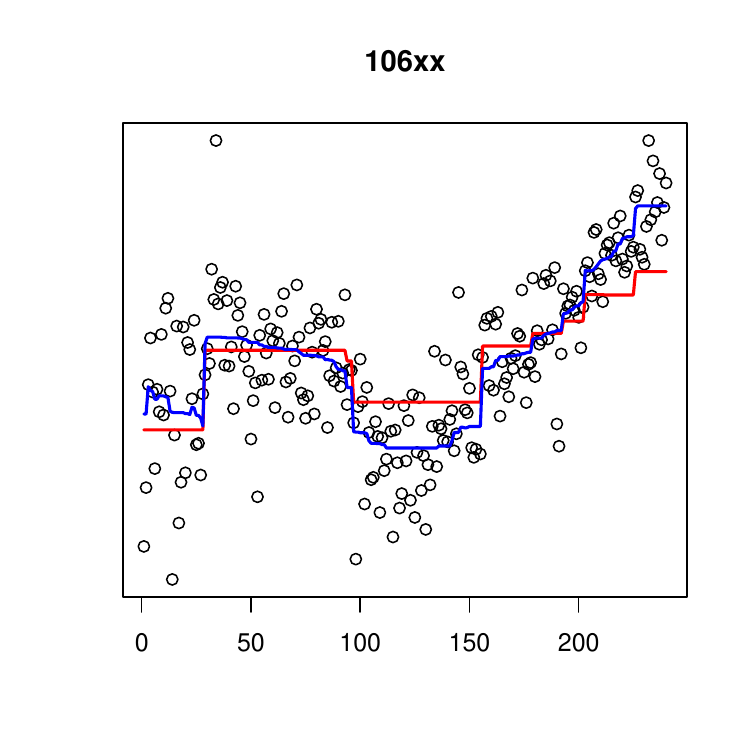}
  \includegraphics[width=0.135\textwidth]{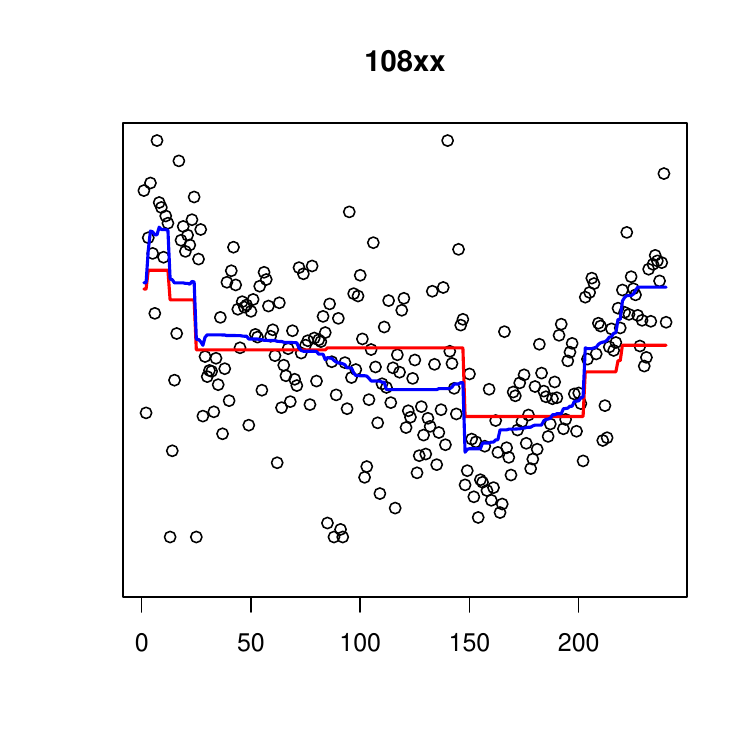}\\
  \includegraphics[width=0.135\textwidth]{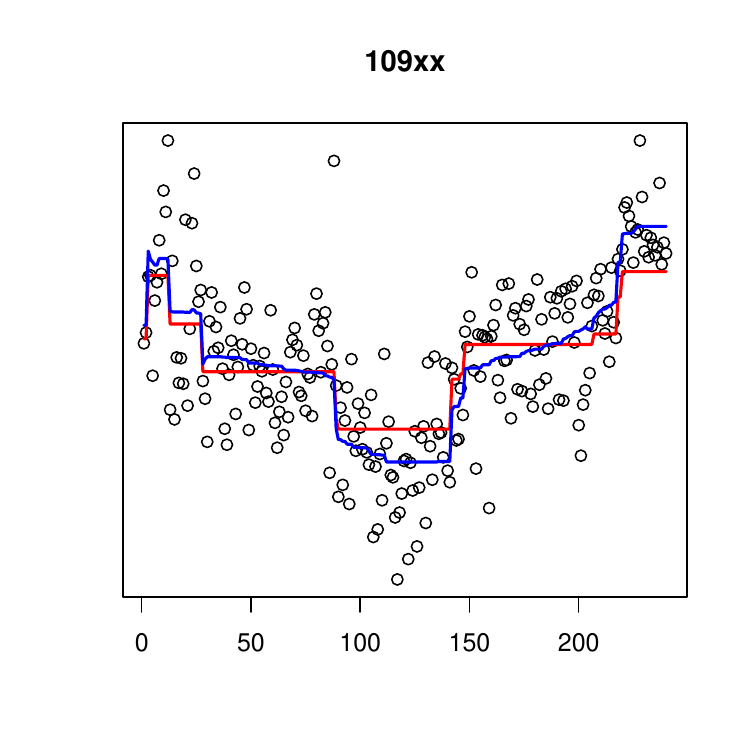}
  \includegraphics[width=0.135\textwidth]{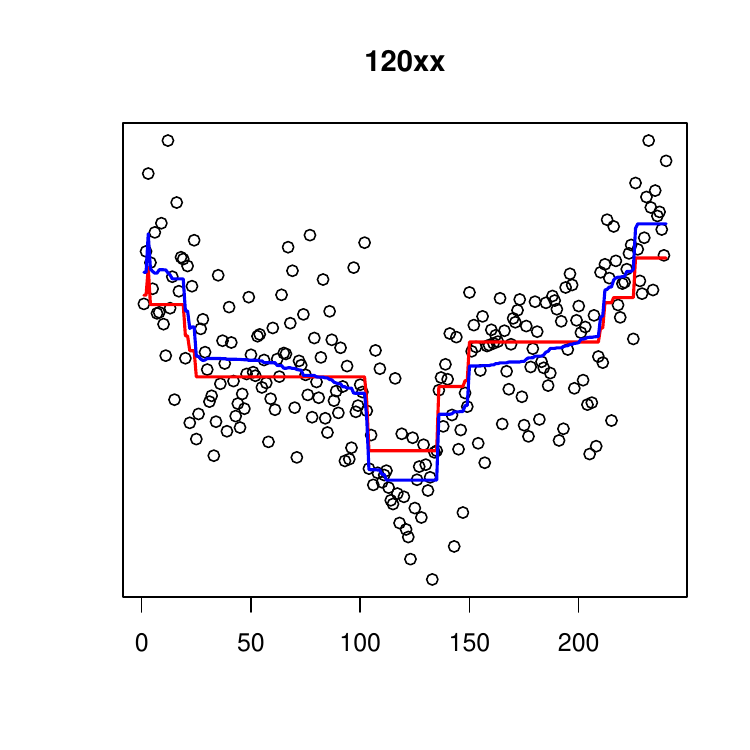}
  \includegraphics[width=0.135\textwidth]{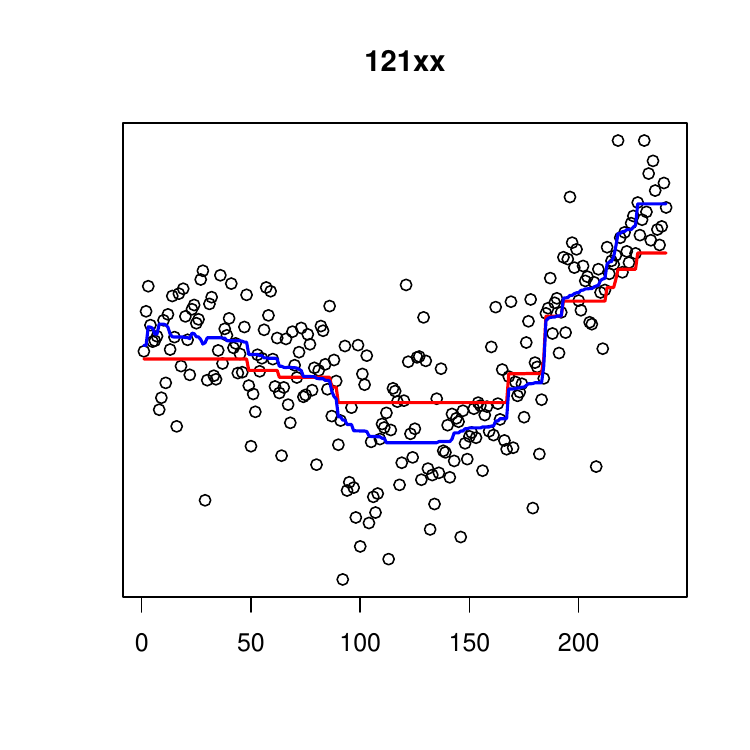}
  \includegraphics[width=0.135\textwidth]{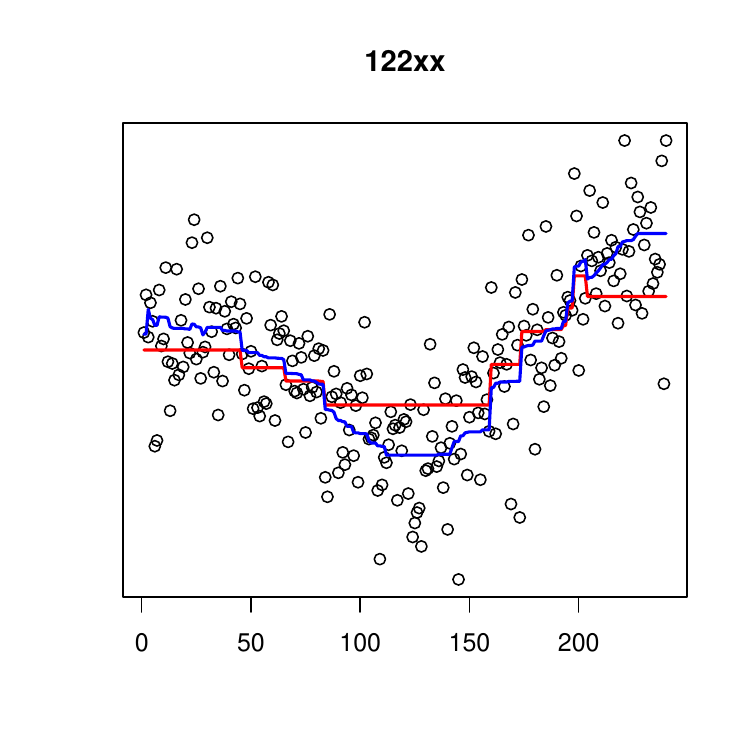}
  \includegraphics[width=0.135\textwidth]{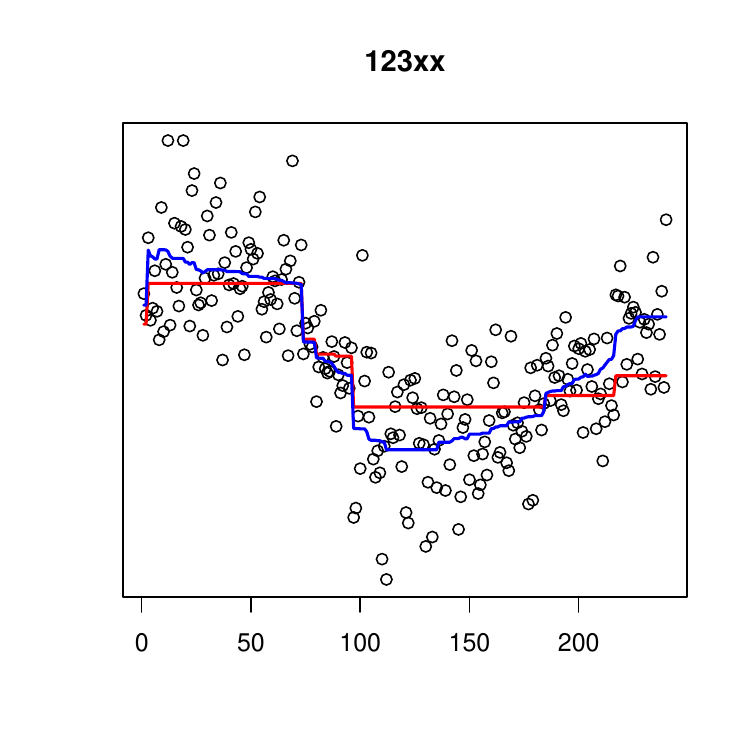}
  \includegraphics[width=0.135\textwidth]{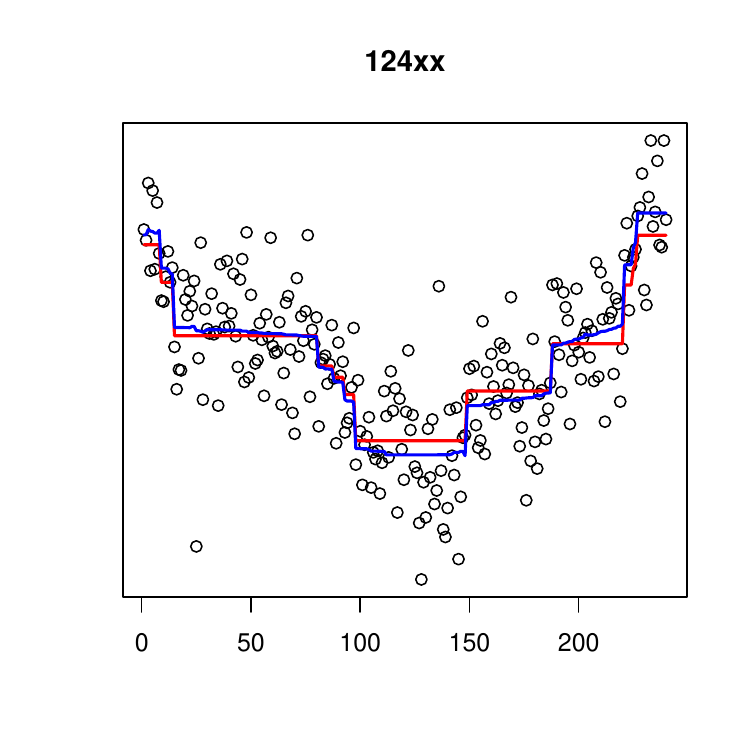}
  \includegraphics[width=0.135\textwidth]{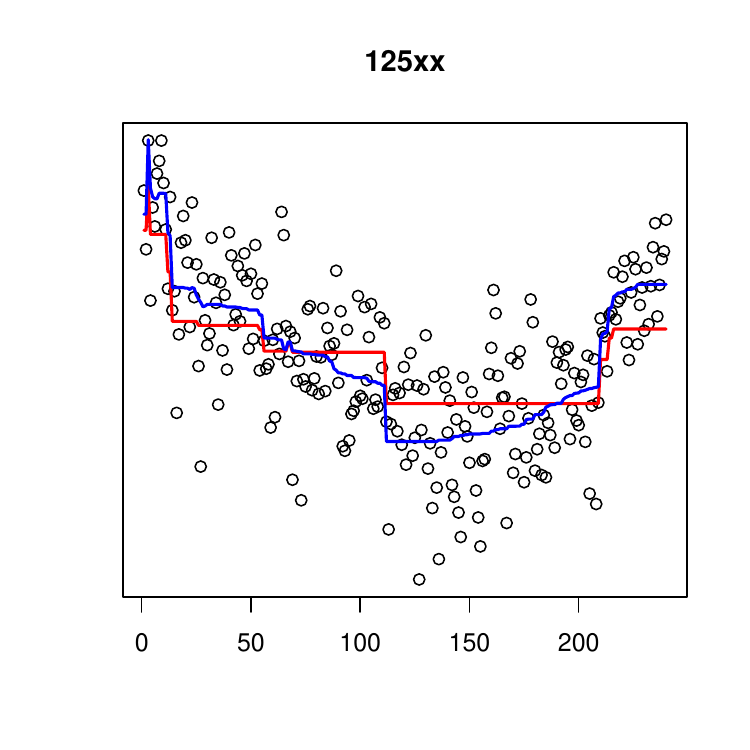}\\
  \includegraphics[width=0.135\textwidth]{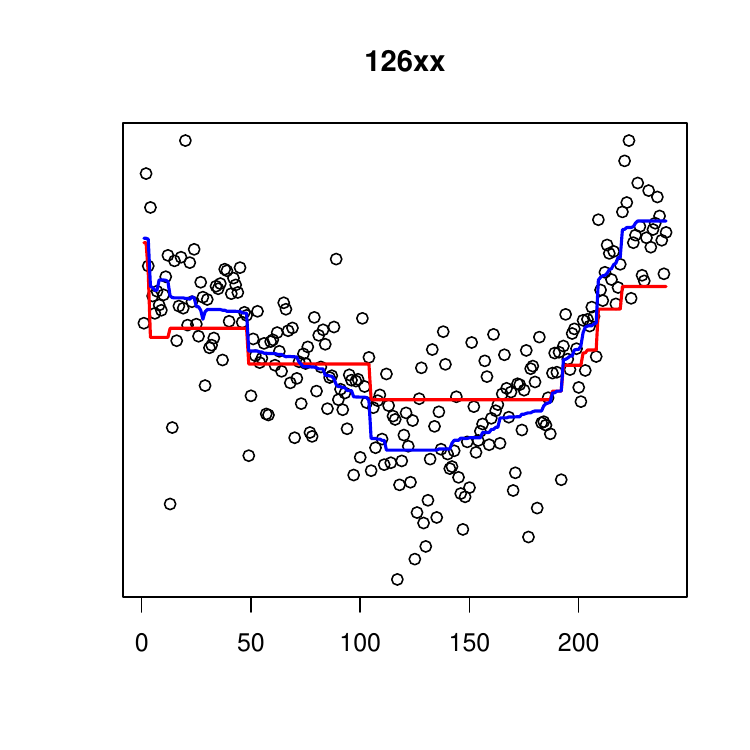}
  \includegraphics[width=0.135\textwidth]{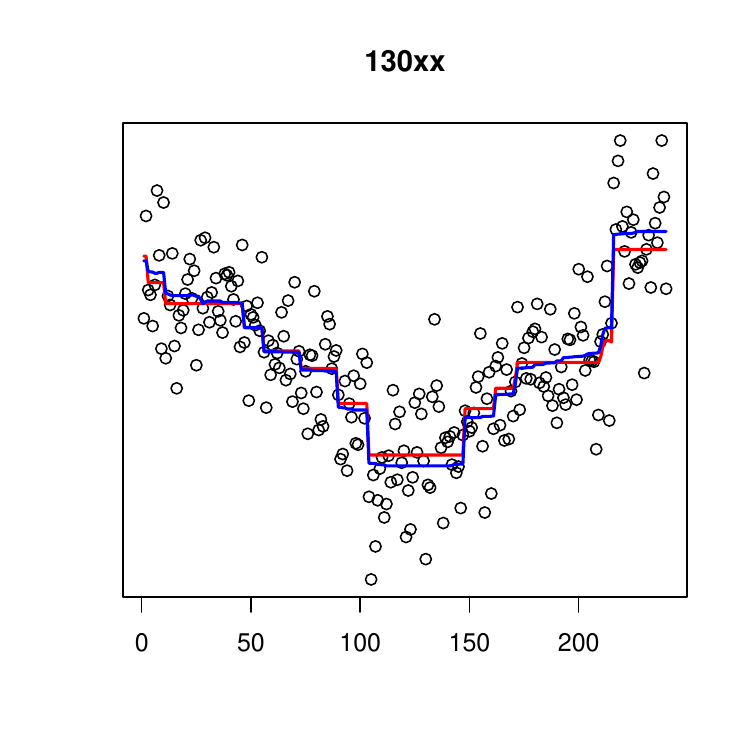}
  \includegraphics[width=0.135\textwidth]{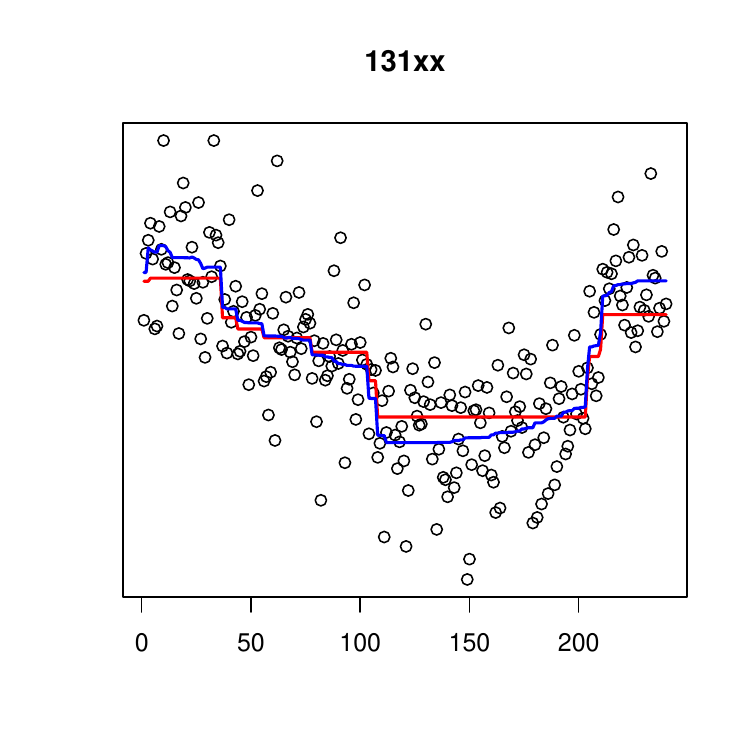}
  \includegraphics[width=0.135\textwidth]{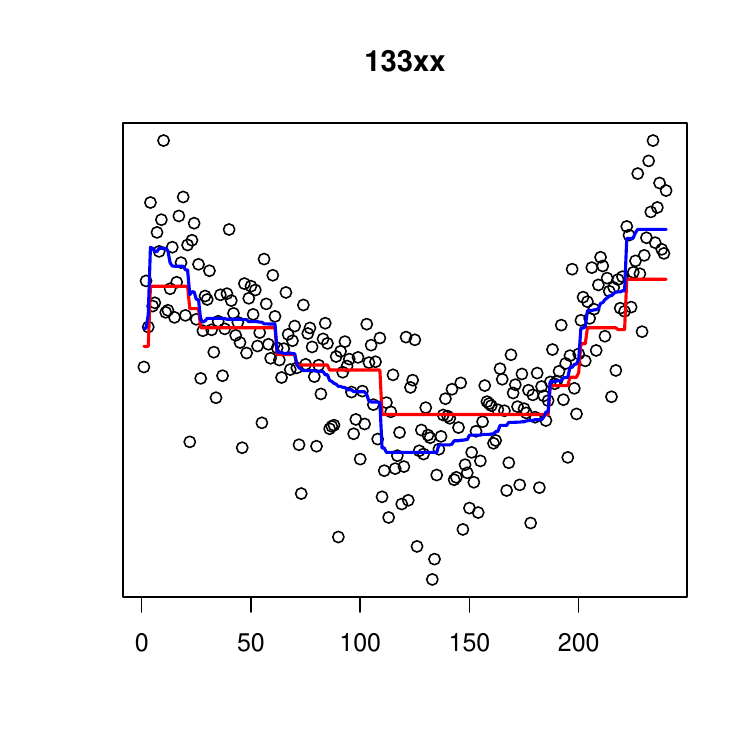}
  \includegraphics[width=0.135\textwidth]{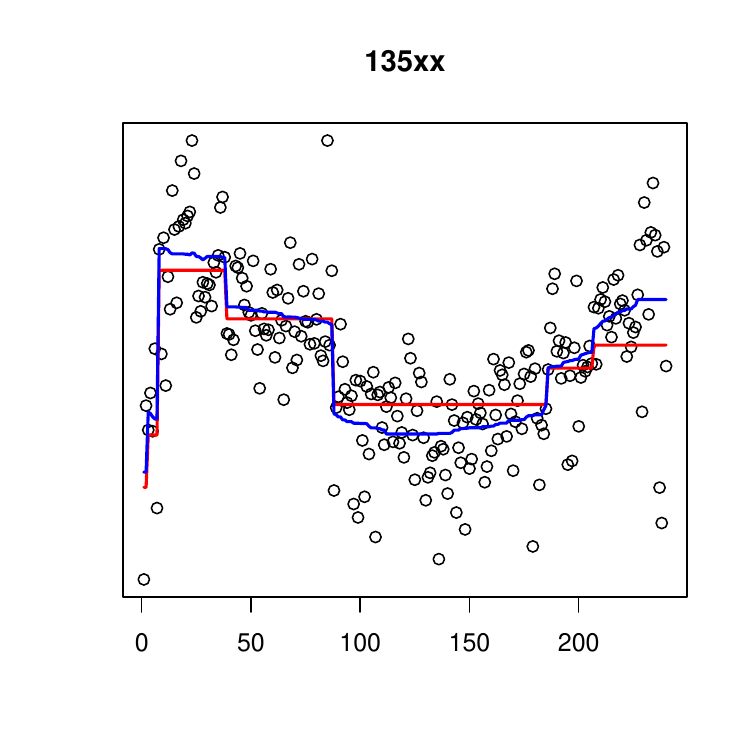}
  \includegraphics[width=0.135\textwidth]{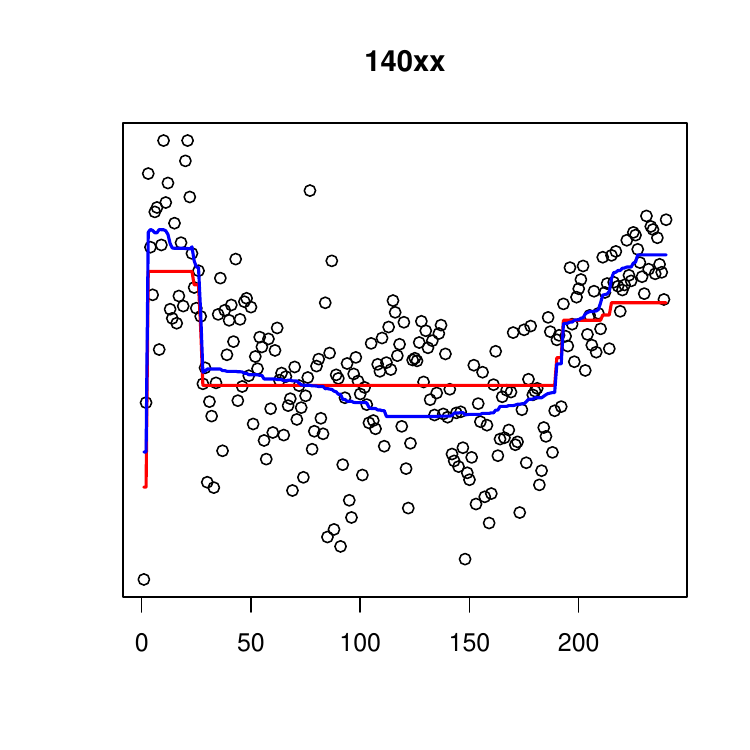}
  \includegraphics[width=0.135\textwidth]{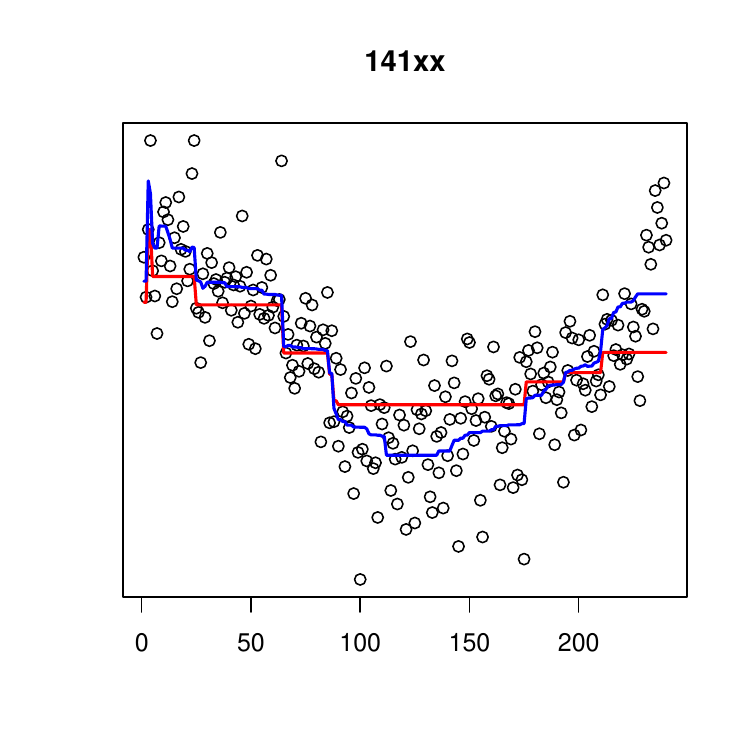}
  \caption{Condominium prices for all considered {postal} code areas in Berlin as time series plots after applying the normal transformation. The solid line (red) corresponds to the estimated time-dependent regional mean level $\hat{\xvec{a}}_t$, and the dashed line (blue) is the estimated overall mean-level {$\hat{\xvec{\alpha}}_t = (\xmat{I} - \hat{\xmat{W}})^{-1} \hat{\xvec{a}}_t$}. In the top row, the {postal} code areas 107xx and 134xx are shown, for which we additionally depict the estimated weights in Figure \ref{fig:Out_Berlin}.}\label{fig:CP_Berlin}
\end{figure}

Figure \ref{fig:CP_Berlin} depicts the {PIT-transformed} real estate prices for all spatial locations as time-series plots. As can be seen, the prices show different patterns depending on the {post} code area. While the mean stays constant over the considered time period for some locations, e.g., 102xx or 103xx, there is a clear decrease and increase of the prices for other locations, such as 120xx, 125xx, or 130xx. In particular, we observe declining prices for the first 8-10 years of our study and rising prices in the most recent 10 years. The estimated coefficients $\hat{\xvec{a}}_t$ are reported in Figure \ref{fig:CP_Berlin} as red lines. {As mentioned earlier, it} is worth noting that these estimated coefficients do not correspond to the estimated {overall mean levels $\hat{\xvec{\alpha}}_t = (\xmat{I} - \hat{\xmat{W}})^{-1} \hat{\xvec{a}}_t$} depicted by the dashed blue curves. Hence, $\hat{\xvec{a}}_t$ should be interpreted as regional baseline prices, or regional mean, without accounting for spillover effects from neighboring areas. Deviations from the overall mean {$\hat{\xvec{\alpha}}_t$} can therefore be associated with higher/lower prices spilling over from areas, which are connected for reasons such as spatial proximity, similar life style, and culture.

\begin{figure}
  \centering
  \includegraphics[width=0.32\textwidth]{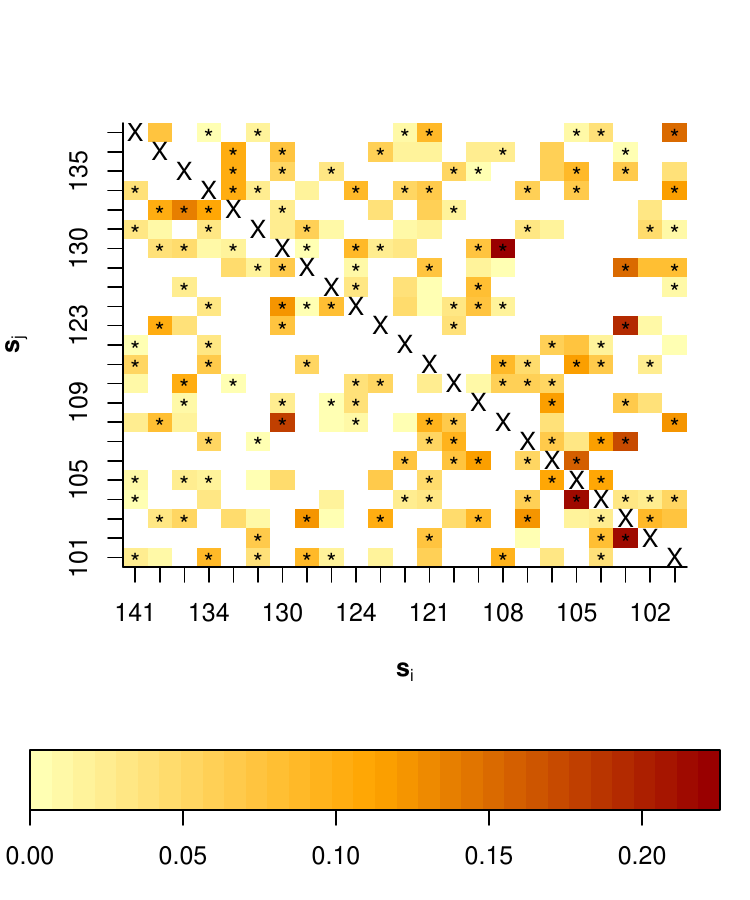}
  \includegraphics[width=0.32\textwidth]{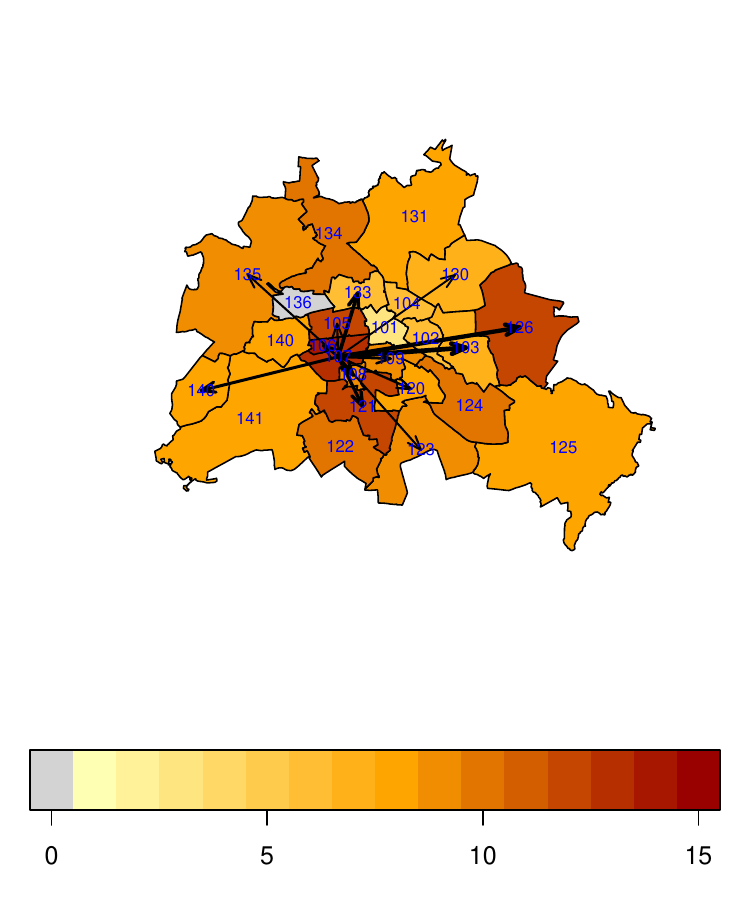}
  \includegraphics[width=0.32\textwidth]{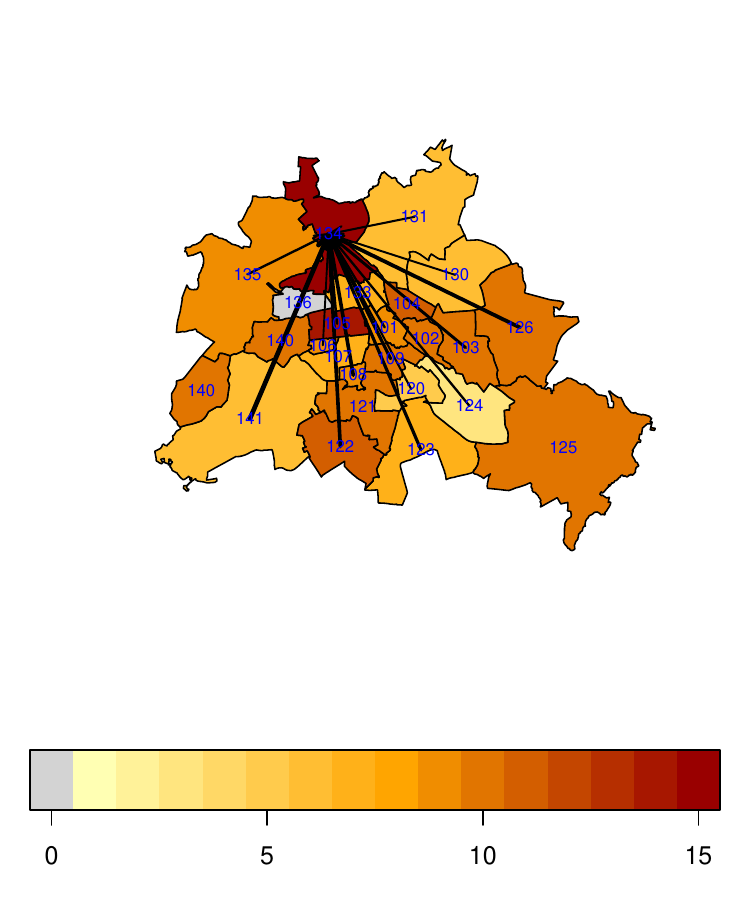}
  \caption{Estimated spatial weighting matrix $\hat{\xmat{W}}$ (left) and number of outgoing (center) and incoming links (right) for each postal code area. In the center, the links originated in area 107xx are exemplarily depicted as arrows. Moreover, a second example of all links influencing area 134xx can be seen in the right-hand map {(note that all arrows point to area 134xx and therefore overlap)}. The line width corresponds to the strength of the spatial dependence. The gray-colored area 136xx was excluded from our analysis.}\label{fig:Out_Berlin}
\end{figure}

Further insights on these connections can be gained from the estimated spatial weighting matrix $\hat{\xmat{W}}$, shown in Figure \ref{fig:Out_Berlin}. The entries of the $i$-th row represent how a location $\xvec{s}_i$ would be influenced by the locations $\xvec{s}_j$, where darker colors represent stronger influences. By contrast, the entries of the $j$-th column show how location $\xvec{s}_j$ influences all other regions. Links estimated to be zero are not colored. Furthermore, we highlight all mutual connections, i.e., $\xvec{s}_i$ influence $\xvec{s}_j$ and vice versa, with an asterisk. These two-way relations are not necessarily equal, meaning that $\hat{w}_{ij}$ {can be different from} $\hat{w}_{ji}$. Moreover, there are several links that are estimated to be directed or one-sided, e.g., 141xx to 140xx (i.e., $\hat{w}_{12} > 0$, but $\hat{w}_{21} = 0$). In order to assess the degree of spatial dependence, we computed the infinity norm of $\hat{\xmat{W}}$, which is equal to $0.404$. Thus, the real-estate prices show a moderate spatial dependence.
% 0.4041586

To illustrate the location of highly influencing or affected areas, we additionally depict the total number of outgoing and incoming links in Figure \ref{fig:Out_Berlin}. We also plot the weights of two selected locations as arrows, with the line width corresponding to the estimated value of the weight. In the central map of Figure \ref{fig:Out_Berlin}, the focus is on region with the {postal} code 107xx, which covers large parts of Berlin-Mitte and Berlin-Charlottenburg. This area is the city center of the former West Berlin and can be seen to influence nearly all urban areas of the city. For instance, there are no links to Berlin-K\"{o}penick (125xx) or Berlin-Pankow (131xx), which are the outer districts with independent quarter centers having their own city halls, central churches, and market places. These districts can be seen as smaller cities located within Berlin. 

The region that was influenced the most by other districts is located in the north of Berlin. It has postal code 134xx and covers mostly the areas of Berlin-Reinickendorf, including the {former} airport Berlin-Tegel (TXL).\footnote{{The airport was closed in 2021, so after the time period covered in this study.}} This region is also characterized by forests and bodies of water, as well as exclusive residential areas. 

{Finally}, it is worth noting that the estimated matrix $\hat{\xmat{W}}$ is not necessarily associated with spatial proximity. {The} first geographic law does {also} not have to be fulfilled (cf. \citealt{Tobler70}){, as we mentioned earlier}. {Hence, it is important to add an geographical interpretation after the estimation, e.g., as we did above.} Moreover, the {studied} geographic area is relatively `homogeneous' and the spatial units are small. Thus, spatial effects dropping off with distance could not be expected, as the covered area is simply not large enough. Thus, {we suggest that} $\hat{\xmat{W}}$ should rather be interpreted as the adjacency matrix of a spatial network.

In summary, we can state that the algorithm captured the structure of the data in terms of locally varying mean levels and spatial dependence. It is important to note that this obtained solution fulfills the identifiability condition given in Proposition \ref{prop:1}. In this way, we gained new insights into spatial relationships that could not be detected by standard distance-based models. Notably, we have shown that the prices of condominiums depend not only on direct neighbors, but also on areas that are farther away. This dependence also does not diminish with increasing distance, which is often assumed to be the case in spatial econometrics. Thus, the proposed modeling approach is able to capture other latent factors, such as socio-economic, cultural, and lifestyle factors. Moreover, the estimated regional/local mean levels $\xvec{a}_t$ show obvious change points and temporal dynamics in the real estate prices. Specifically, the temporal pattern differs between the locations, such that the changes do not necessarily occur at the same time points for each area.

\section{Conclusion}\label{sec:conclusion}

A common and well-known issue in spatial econometrics is the need to find a suitable spatial weighting matrix. This matrix defines the structure of the spatial dependence and is usually unknown when applying the model to real data. Although one might gain insight into the spatial dependence structure using tools like the spatial variogram, irregular dependencies, which are not a function of the differences between two locations $\xvec{s}_i - \xvec{s}_j$ only, would still be undetected. Indeed, due to the curse of dimensionality, the spatial weights are typically not estimated, but instead replaced by predefined weights multiplied with unknown scalar(s) to be estimated. {Thus, it is essential to interpret all results conditional on the choice of these weights.}

We addressed this important issue in spatial econometrics by proposing a penalized regression approach that allowed us to estimate the entire spatial weights matrix. In addition, we accounted for possible structural breaks within the panel data, which affected both the mean level of the location and where the change point occurred, as well as mean level of all other locations. In future research, it is important to also consider more complex models. For instance, we excluded potential influences from exogenous regressors {or temporal autoregressive effects}. That is, omitted exogenous variation would be picked up by the locally varying mean levels and could be erroneously mistaken as change points. {Likewise, missing sources of cross-correlation are reflected into the estimated spatial weights, because an independent error process is assumed. Thus,} another important generalization of the method would be to allow for more complex dependence structures of the error term, like spatial autoregressive or moving average errors. In addition, spatially correlated regressors as well as multivariate spatial models would be important extensions from a practical point of view.

The estimation procedure proposed in this paper is a two-stage approach in which the dimension of the parameter space is reduced in the first stage. In a second stage, the full model, including the spatial dependence and the changes in the mean level, is estimated. To obtain the parameter estimates, a constrained LASSO approach has been implemented and {a} completely novel criterion to select the penalty parameter has been proposed. More precisely, this new criterion is based on the distance between the sample autocorrelation in space and the spatial autocorrelation implied by the estimated model. Since the local mean levels and the spatial dependence are strongly interconnected\footnote{{Among all considered settings in the simulations, we have seen that} if the mean levels are correctly estimated, so is the spatial dependence and vice versa. By contrast, wrongly estimated spatial links can be captured by the local mean levels, which would also differ from the true values in this case, and vice versa.} {and the mean levels are penalized in the first step}, it {seems to be} sufficient to solely rely on this spatial measure. Up to now, we only focused on spatial econometrics models, but there is also the broad field of geostatistics using mostly covariance-based models. For these models, the spatial dependence is directly modeled by the $n\times n$-dimensional covariance matrix and a covariance function. In the future, it would be interesting to see whether a similar LASSO-type approach can be used to estimate the covariance matrix directly, while still having the possibility to construct a covariance function for kriging.

We analyzed the performance of our approach via an extensive simulation study. Generally, the procedure works well for small, medium, and large spatial dependence. Both sensitivity and specificity for detecting spatially dependent locations are reasonably large, and changes at different time points for different locations can be detected. However, the estimation procedure tends to estimate symmetric spatial dependence structures, where the respective weights are not necessarily equal. More precisely, this meant that the symmetry in the positive entries, i.e., if a weight $w_{ij} > 0$, then $w_{ji}$ should also be positive. Due to these issue with identifying the direction of the dependence, the approach works better for data in which there is this type of symmetric spatial dependence. Thus, if the weights are assumed to be symmetric a priori, there are fewer free parameters and the spatial dependence can easier be {identified}. {Moreover, we provide theoretical results for the identifiability of the parameters in Proposition \ref{prop:1}.} {Note that the restriction to symmetric weights} would only lead to better results, if the true but unknown spatial dependence is symmetric in fact. {If} the dependence is not symmetric, one should be aware that weaker entries {will} spuriously occur on both sides. Thus, we recommend to estimate the full spatial weights matrix in any case. In general, the individual weights should be interpreted with caution, because the problem is high-dimensional and there is not necessarily a unique solution of the constrained LASSO estimator.

Finally, we illustrated the proposed two-stage LASSO approach through an empirical example of Berlin real estate prices. On the one hand, we can see that the spatial dependence differs from classically applied dependence schemes, such as contiguity matrices. In particular, the postal code areas, which we considered to be spatial units, were not only dependent on direct neighbors, but also on areas that were farther away but had the possibility of good public transport connections and similar lifestyles, among other factors. On the other hand, the mean level of the prices changed over time, as modeled by the change points. The {local} means, which should be interpreted as prices without accounting for prices in the neighboring areas, were found to be different for all spatial locations. For some areas, the price level was more or less constant over time; for others, we observe declining prices in the first period of our study and rising prices in the second one. Moreover, it is possible to distinguish between {local} mean levels and overall mean levels, which meant that one can see if natural prices increased or decreased due to the differing price levels in surrounding areas.

\section*{Acknowledgment}

{We want to thank the two anonymous reviewers of JCGS whose suggestions and comments helped us a lot in the revision process. We would also like to thank the Associate Editor for his detailed and thorough comments.}

% \bibliography{Bib1}

\newpage

\begin{appendix}

\section*{Proof of Proposition \ref{prop:1}}

\begin{proof}[Proof of  Proposition \ref{prop:1}]
Let $\xvec{Y}_L = (\xvec{Y}_{1 \cdot}, \ldots, \xvec{Y}_{T \cdot})'$, $\xvec{\varepsilon}_L = (\xvec{\varepsilon}_{1}, \ldots, \xvec{\varepsilon}_{T})'$ and $\xvec{a}_L = (\xvec{a}_{1}, \ldots, \xvec{a}_{T})'$ {be stacked $nT$-dimensional vectors in the long format. Moreover, let $\vartheta = (\vartheta_1, \ldots, \vartheta_p)'$ be the $p$-dimensional parameter of all structural parameters given by the active sets $\mathcal{A}_a$ and $\mathcal{A}_w$, that is $p = |\mathcal{A}_a \cup \mathcal{A}_w|$.} From \eqref{eq:model2}, we get the stacked, reduced form of the model
\begin{equation}\label{eq:reduced_form_full}
  \xvec{Y}_L = \xvec{H}(\vartheta) + \xvec{\nu} =  \xmat{I}_T \otimes (\xmat{I}_n - \xmat{W})^{-1} \xvec{a}_L + \xvec{\nu}
\end{equation}
with $\otimes$ being the Kronecker product, $\xmat{I}_d$ the $d$-dimensional identity matrix, and
\begin{equation}
 \xvec{\nu} = \xmat{I}_T \otimes (\xmat{I}_n - \xmat{W})^{-1} \xvec{\varepsilon}_L
\end{equation}
Thus, Theorem 6 of \cite{Rothenberg71} can be applied and we have to show that the rank of {$\tilde{\xmat{H}} = \left( \frac{\partial H_k(\vartheta)}{\partial \vartheta_j} \right)_{\substack{k= 1, \ldots, nT \\ j = 1, \ldots, p}}$ is equal to $p$.} Differentiating with respect to the local mean levels, we get
\begin{equation}
    \left(\frac{\partial H_k(\vartheta)}{\partial \xvec{a}_L }\right)_{k = 1, ..., nT} = \xmat{I}_T \otimes \xmat{S} \, ,
\end{equation}
which is independent of $t$ (i.e., it has a block diagonal structure). Furthermore,
\begin{equation}
  \left(\frac{\partial H_k(\vartheta)}{\partial w_{ij}}\right)_{k = 1, ..., nT} = (\xmat{I}_T \otimes \xmat{S} \xvec{\iota}_{ij} \xmat{S}) \xvec{a}_L \, .
\end{equation}
Since all columns of $\left[ \left(\xmat{I}_T \otimes \xmat{S} \right)_{\mathcal{A}_a}, \left( \left((\xmat{I}_T \otimes \xmat{S} \xvec{\iota}_{ij} \xmat{S})\xvec{a}_L \right)_{\substack{i,j = 1, \ldots, n \\ i \neq j}}\right)_{\mathcal{A}_w} \right]$ are assumed to be linearly independent and the number of parameters is smaller than or equal to $nT$, $\tilde{\xmat{H}}$ has {full rank. This proves the parameter identifiability.}
\end{proof}

\begin{proof}[Proof of Proposition \ref{prop:first_stage}]
Suppose that $\alpha_{t,i} = \alpha_{t-1,i}$ for all $i = 1, \ldots, n$. Hence,
\begin{equation*}
  \xvec{\alpha}_{t} = (\xmat{I} - \xmat{W})^{-1} \xvec{a}_t = (\xmat{I} - \xmat{W})^{-1} \xvec{a}_{t-1} = \xvec{\alpha}_{t-1}
\end{equation*}
and
\begin{equation*}
  (\xmat{I} - \xmat{W})^{-1} (\xvec{a}_t - \xvec{a}_{t-1}) = \xvec{0} \, ,
\end{equation*}
if and only if $\xvec{a}_t = \xvec{a}_{t-1}$. Since $\varrho(\xmat{W}) < 1$, this is an immediate consequence of the rank-nullity theorem (e.g. \citealt{Banerjee14_algebra}).
\end{proof}

\begin{figure}
	\centering
	\input{W_Example_3_3.tex}
	\caption{Row-standardized Queen's contiguity matrix used for the simulation in Figure \ref{fig:example}. }\label{fig:W_Example}
\end{figure}

\end{appendix}

\end{document}